\documentclass[letterpaper,usenatbib]{mn2e}
\usepackage{amsmath}
\usepackage[breaklinks, colorlinks, citecolor=blue]{hyperref}
\usepackage{epsfig}
\usepackage{multirow}
\usepackage{times}
\usepackage{verbatim}
\def\gtsim{~\rlap{$>$}{\lower 1.0ex\hbox{$\sim$}}}
\def\ltsim{~\rlap{$<$}{\lower 1.0ex\hbox{$\sim$}}}

\title[NGC 4945 as observed by ALMA]{Free-free and H42$\alpha$ emission from the dusty starburst within NGC 4945 as observed by ALMA} 

\author[G. J. Bendo et al.]
    {G. J. Bendo$^{1,2}$, C. Henkel$^{3,4}$, M. J. D'Cruze$^1$, C. Dickinson$^1$, G. A. Fuller$^{1,2}$, A. Karim$^5$ \newauthor \\
    $^1$   Jodrell Bank Centre for Astrophysics,
           School of Physics and Astronomy, The University of Manchester, 
           Oxford Road, Manchester M13 9PL,\\ United Kingdom\\
    $^2$   UK ALMA Regional Centre Node\\
    $^3$   Max-Planck-Institut f\"ur Radioastronomie, Auf dem H\"ugel 69, 53121 
           Bonn, Germany\\
    $^4$   Astronomy Department, Faculty of Science, King Abdulaziz University, 
           P.O. Box 80203, Jeddah 21589, Saudi Arabia\\
    $^5$   Argelander-Institut f\"ur Astronomie, Universit\"at Bonn, Auf 
           dem H\"ugel 71, D-53121 Bonn, Germany \\
}
\date{}
\pagerange{\pageref{firstpage}--\pageref{lastpage}}
\pubyear{}

\begin{document}
\label{firstpage}
\maketitle

\begin{abstract}
We present observations of the 85.69~GHz continuum emission and H42$\alpha$ line emission from the central 30~arcsec within NGC~4945.  Both sources of emission originate from nearly identical structures that can be modelled as exponential discs with a scale length of $\sim$2.1~arcsec (or $\sim$40 pc).  An analysis of the spectral energy distribution based on combining these data with archival data imply that 84\%$\pm$10\% of the 85.69~GHz continuum emission originates from free-free emission.  The electron temperature is 5400$\pm$600~K, which is comparable to what has been measured near the centre of the Milky Way Galaxy.  The star formation rate (SFR) based on the H42$\alpha$ and 85.69~GHz free-free emission (and using a distance of 3.8~Mpc) is 4.35$\pm$0.25 M$_\odot$ yr$^{-1}$.  This is consistent with the SFR from the total infrared flux and with previous measurements based on recombination line emission, and it is within a factor of $\sim$2 of SFRs derived from radio data.  The {\it Spitzer} Space Telescope 24~$\mu$m data and Wide-field Infrared Survey Explorer 22~$\mu$m data yield SFRs $\sim$10$\times$ lower than the ALMA measurements, most likely because the mid-infrared data are strongly affected by dust attenuation equivalent to $A_V=150$.  These results indicate that SFRs based on mid-infrared emission may be highly inaccurate for dusty, compact circumnuclear starbursts.
\end{abstract}


\section{Introduction}
\label{}
\addtocounter{footnote}{5}

The Atacama Large Millimeter/submillimeter Array (ALMA) is capable of detecting two different forms of emission from photoionized gas in the star forming regions within other galaxies.  First, ALMA can measure continuum emission at 85-100~GHz where the spectral energy distributions (SEDs) of galaxies are dominated by free-free emission \citep[e.g.][]{peel11}.  Second, ALMA is sensitive enough to detect recombination line emission that appears at millimetre and submillimetre wavelengths.  Both free-free and millimetre recombination line emission as star formation tracers have advantages over ultraviolet, optical, and near-infrared tracers in that they are unaffected by dust attenuation.  Unlike infrared or radio continuum emission, the millimetre continuum and recombination line emission directly traces of photoionized gas and therefore should be more reliable for measuring accurate star formation rates (SFRs).  For additional discussion about this, see \citet{murphy11}.

Millimetre continuum observations of nearby galaxies have been relatively straightforward, but the  recombination line emission has been more difficult to detect.   Before ALMA, the millimetre recombination lines have been detected in multiple star forming regions within the Milky Way \citep[e.g. ][]{waltman73, wilson84, gordon89, gordon90}, but extragalactic millimeter recombination line emission has only been detected in M82 \citep{seaquist94, seaquist96}, NGC 253 \citep{puxley97}, and Arp 220 \citep{anantharamaiah00}.  ALMA is capable of reaching sensitivity levels at least an order of magnitude better than other telescopes \citep[see][for a technical overview]{remijan15} and can therefore lead to detections in many more nearby infrared-luminous sources than was previously possible \citep{scoville13}.  At this time, however, ALMA detections of specifically recombination line emission have been limited.  \citet{bendo15b} and \citet{meier15} reported the detection of millimetre recombination line emission from the nearby starburst galaxy NGC~253, and \citet{bendo15b} used the 99.02~GHz continuum and H40$\alpha$ (99.02 GHz) line emission to measure electron temperatures, SFRs, and (along with near-infrared data) dust attenuation in the centre of the galaxy.  \citet{scoville15} reported the marginal detection of H26$\alpha$ emission from Arp 220, but the presence of the nearby HCN(4-3) line made it difficult to perform any analysis with this line detection.

In this paper, we report on the detection of 85.69~GHz and H42$\alpha$ recombination line emission from the centre of NGC~4945, a nearby \citep[3.8$\pm$0.3~Mpc;][]{karachentsev07, mould08} spiral galaxy with an optical disc of 20.0$\times$3.8~arcmin \citet{devaucouleurs91}.  While the galaxy contains a composite active galactic nucleus and starburst nucleus and while the AGN is one of the brightest hard X-ray source as seen from Earth \citep{done96}, analyses based on near- and mid-infrared emission lines imply that the starburst is the primary energy source for exciting the photoionized gas and that the AGN is heavily obscured \citep{marconi00, spoon00, spoon03, perezbeaupuits11}.  The central region is very dusty, so even near-infrared star formation tracers such as Pa$\alpha$ lines are strongly attenuated \citep{marconi00}.  In such a source, millimetre star formation tracers would provide a much more accurate measurement of SFR.

Previously, the only detection of millimetre or radio recombination line emission from NGC~4945 were the H91$\alpha$ and H92$\alpha$ lines observed by \citet{roy10}.  However, these low-frequency lines are significantly more affected by masing and collisional broadening effects than millimetre lines \citep{gordon90}, and the photoionized gas is typically optically thick at radio wavelengths but optically thin at millimetre wavelengths.  Given this, the millimetre continuum and line data should provide a better measure of SFR than the radio line data.

After a description of the data in Section~\ref{s_data}, we present the analysis and the discussion of the results in five sections.  Section~\ref{s_image} presents the images and also shows simple models fit to the data. Section~\ref{s_spec} shows the spectra as well as maps of the parameters describing the recombination line profiles. Section~\ref{s_sed} presents the SED of the emission from the central source between 3 and 350~GHz, which we used to determine the fraction of continuum emission originating from free-free emission.  Section~\ref{s_te} presents the derivation of electron temperatures ($T_e$) from the ratio of the millimetre recombination line to the free-free continuum emission and discusses the results in the context of other measurements within NGC 4945, the Milky Way Galaxy, and other galaxies.  Section~\ref{s_sfr} describes the measurement of SFRs from the millimetre data, the measurement of SFRs from infrared data, and SFR measurements based on published radio continuum photometry as well as previously-published SFR measurements based on other radio data.  We then perform a comparison of these various SFR measurements to identify consistencies between the measurements and to understand the reasons why some measurements may appear inconsistent with others.  A summary of all of the results is presented in Section~\ref{s_conclu}.

\section{Data}
\label{s_data}

\subsection{ALMA data}
\label{s_data_alma}

The ALMA observations, which were part of project 2012.1.00912.S (PI: C. Henkel), were performed in a single execution block on 24 January 2013 with 31 operational antennas (after flagging).  The array was in a compact configuration with baselines extending from 15 to 445~m.  The observations consist of a single pointing aimed at the centre of the galaxy, and the total on-source integration time was 33.94 min.  The spectral window containing the H42$\alpha$ line was centered on 85.895 GHz, contained 1920 channels with a width of 976.6 kHz (3.4 km s$^{-1}$), and included both polarizations.  The bandpass calibrator was J1427-4207, the flux calibrator was J1107-448, and the phase calibrator was J1248-4559.

The data were reprocessed using the Common Astronomy Software Applications ({\sc CASA}) version 4.5.0.  First, we applied a series of a priori calibration steps, including phase corrections based on water vapour radiometer measurements and amplitude corrections based on system temperatures, to the visibility data.  Next, we flagged noisy or anomalous visibility data.  After this, we derived and applied corrections to the phase and amplitude as a function of frequency and time.  The visibility data for the spectral window containing the H42$\alpha$ line were then converted into image cubes using {\sc clean} in {\sc CASA} with the Cotton-Schwab imager algorithm and with natural weighting.  Two versions of the image cube were created.  One version of the image cube included the continuum emission.  The other version was continuum-subtracted based on a linear fit to line-free regions in visibility channels 400-560 (sky frequencies 86.288-86.444~GHz), 1000-1200 (sky frequencies 85.663-85.859~GHz), and 1440-1500 (sky frequencies 85.370-85.429~GHz); no other regions in the spectrum were clear of spectral lines.  The final image cubes cover the full frequency range of the spectral window (sky frequencies 84.960-86.831~GHz), and the image cube channels are set to be 4 times the size of the visibility channels, which results in a channel width of 3.906~MHz (13.7~km~s$^{-1}$).  The final image cubes have pixel scales of 0.5~arcsec and dimensions of 256$\times$256 pixels (128$\times$128~arcsec), which covers the primary beam.  The full-width at half-maximum (FWHM) of the reconstructed beams are 2.6$\times$2.3~arcsec.  Primary beam corrections are applied to the images after they are created.  The total H42$\alpha$ flux in each map pixel was calculated by summing the signal between channels 310 and 365 (sky frequencies 85.401-85.616~GHz) in the continuum-subtracted image.   The continuum in each map pixel was calculated by subtracting the cube without the continuum emission from the cube with the continuum and then finding the median of all continuum values between channels 310 and 365.  

The accuracy of the measured SFRs will depend upon the accuracy of the flux calibration based on J1107-448.  We used all archival flux monitoring data from between 120 days before to 120 days after the observations to estimate the flux densities for J1107-448.  In the spectral window with the H42$\alpha$ line, the flux density at the central frequency of 85.895~GHz is estimated as $1.44\pm0.03$~Jy.  However, the measurements in band 3 vary by 6\% over this time period, so we will use that as the calibration uncertainty for both the line and continuum measurements.  As an additional check of the flux calibration, we compared the flux densities from a continuum image of J1427-4207 to archival flux monitoring data \footnote{J1248-4559 was not observed 300 days before or after the NGC 4945 observations and therefore could not be used in this check.}.  The flux density at this frequency estimated from the archival data is 7.5$\pm$0.2~Jy, but the archival data for band 3 vary by 17\% (or 1.2~Jy) within the 240 day period we used.  The flux density we measure at 85.895~GHz is 7.99~Jy, which is consistent with this measurement to within $\sim$6\%.

Global flux densities for NGC~4945 in the 80-90~GHz range have been measured by the {\it Wilkinson} Microwave Anisotropy Probe and by the {\it Planck} mission.  However, the angular resolution of the data from these telescopes is so broad that it is difficult to extract flux densities for the central starburst for comparison to the ALMA data.  The only published continuum data from the 80-90~GHz range are the 89~GHz continuum data from the Australia Compact Telescope Array (ATCA) published by \citet{cunningham05}.  They did not publish an integrated continuum flux density for the central region but instead only report the results of a Gaussian fit to the central source; the best fitting parameters give a flux density of 0.28~Jy.  The total continuum emission we measure with ALMA at 85.69~GHz is 0.36$\pm$0.02 ~Jy (see Section~\ref{s_sed}).  The numbers differ by $\sim$25\%.  \citet{cunningham05} did not publish any uncertainties related to their measurement, but the typical flux calibration uncertainties for ATCA at the time of the observations was $\sim$30\% (Stevens, private communication).  Given this, the difference does not seem significant.  Additionally, the \citet{cunningham05} data should be treated as a lower limit for the actual flux density since their continuum image shows some non-Gaussian asymmetries and extended structure that may not be included in the part of the data described by the Gaussian fit.   Given this, we will assume that the ALMA flux calibration is reliable to within 6\% for our analysis.

\subsection{Infrared data}
\label{s_data_ir}

\begin{table*}
\centering
\begin{minipage}{176mm}
\caption{Technical data for infrared images.}
\label{t_irdata}
\begin{tabular}{@{}cccccccl@{}}
\hline
Telescope &
  Instrument &
  Wavelength &
  Frequency &
  Beam FWHM &
  Flux Calibration &
  Final Map Pixel &
  References for technical information\\
&
  &
  ($\mu$m) &
  (GHz) &
  (arcsec) &
  Uncertainty &
  Size (arcsec) &
  \\
\hline
{\it Spitzer} &
  MIPS &
  24 &
  12500 &
  6 &
  4\% &
  1.5 &
  \citet{engelbracht07} \\
WISE &
  &
  22 &
  13636 &
  17$^a$ &
  2\% &
  1.375 &
  \citet{cutri13}$^b$\\
{\it Herschel} &
  PACS &
  70 &
  4286 &
  5.6 &
  5\% &
  1.6 &
  \citet{altieri13}$^c$ \\
{\it Herschel} &
  PACS &
  100 &
  3000 &
  6.8 &
  5\% &
  1.6 &
  \citet{altieri13}$^c$ \\
{\it Herschel} &
  PACS &
  160 &
  1875 &
  11.4 &
  5\% &
  3.2 &
  \citet{altieri13}$^c$ \\
{\it Herschel} &
  SPIRE &
  250 &
  1200 &
  17.6$^d$ &
  4\% &
  6 &
  \citet{bendo13}, \citet{valtchanov14}$^e$ \\
{\it Herschel} &
  SPIRE &
  350 &
  857 &
  23.9$^d$ &
  4\% &
  10 &
  \citet{bendo13}, \citet{valtchanov14}$^e$ \\
{\it Herschel} &
  SPIRE &
  500 &
  600 &
  35.2$^d$ &
  4\% &
  14 &
  \citet{bendo13}, \citet{valtchanov14}$^e$ \\
\hline
\end{tabular}
$^a$ The observed beam FWHM is 12~arcsec, although the FWHM in the final atlas tiles is 17~arcsec.\\
$^b$ This reference is available from http://wise2.ipac.caltech.edu/docs/release/allwise/expsup/index.html .\\
$^c$ This reference is available from http://herschel.esac.esa.int/Docs/PACS/pdf/pacs\_om.pdf .\\
$^d$ These SPIRE beam FWHM apply to the timeline data, which was used for photometry measurements.  The beam appears broader in the map data.\\
$^e$ This reference is available from http://herschel.esac.esa.int/Docs/SPIRE/spire\_handbook.pdf .\\
\end{minipage}
\end{table*}

For comparison to the ALMA-based SFRs, we calculated SFR using 24~$\mu$m data from the {\it Spitzer} Space Telescope \citep{werner04}, 22~$\mu$m data from the Wide-field Infrared Survey Explorer \citep[WISE; ][]{wright10}, and 70-500~$\mu$m data from the {\it Herschel} Space Observatory \citep{pilbratt10}.  Table~\ref{t_irdata} lists some of the technical details (beam FWHM, flux calibration uncertainties, and final map pixel sizes) for the data we used from each of these telescopes.

The dust emission in the mid-infrared bands originates from a combination of stochastically-heated small dust grains and large dust grains at constant temperatures of $\sim$100~K.  However, the mid-infrared flux at 22 or 24~$\mu$m correlates well with the total infrared luminosity \citep[e.g.][]{rieke09, boquien10, galametz13}, which itself is correlated to star formation if the newly formed stars are the dominant luminosity source.  Emission in the 22 or 24~$\mu$m bands is well-correlated with other star formation tracers \citep[e.g.][]{calzetti05, calzetti07, prescott07, bendo15a}, and multiple relations have been derived for converting the flux densities in these bands to SFRs, as reviewed by \citet{calzetti10} and \citet{lee13}.  Moreover, the spatial resolution of the {\it Spitzer} 24~$\mu$m and WISE 22~$\mu$m data make it easier to separate emission from the central source and the extended disc, whereas this is not as easily done in longer wavelength data from {\it Spitzer} or {\it Herschel} .  Normally, {\it Spitzer} 24~$\mu$m data would be sufficient for tracing hot dust emission, but the {\it Spitzer} 24~$\mu$m data for this galaxy suffer some nonlinearity or saturation effects, as discussed below, which is why the WISE 22~$\mu$m data are also used.

We also include the far-infrared {\it Herschel} data in our analysis because the data, when combined with mid-infrared data, directly trace the bolometric luminosity of star forming regions, particularly regions that are optically thick at optical and near-infrared wavelengths like the centre of NGC~4945.  Also, the relation between 22 or 24~$\mu$m emission and total infrared flux may become nonlinear in infrared-luminous systems \citet[e.g.][]{rieke09}, in which case the total infrared flux may yield different SFRs.  The disadvantage in using the {\it Herschel} data is that the central starburst is more poorly resolved, and emission from the central starburst may be blended with emission from diffuse dust or other star forming regions outside the centre.  None the less, infrared flux densities measured with {\it Herschel} can still be treated as upper limits (although we argue later that the extended disc emission is negligible relative to the emission from the central starburst in this specific galaxy).

\subsubsection{Spitzer 24~$\mu$m data}

The {\it Spitzer} 24~$\mu$m data were acquired with the Multiband Imaging Photometer for {\it Spitzer} \citep[MIPS; ][]{rieke04} in program 40410 (PI: G. Rieke).  The observations were performed in astronomical observation requests 22002432 (executed on 16 July 2007) and 22002688 (executed on 24 July 2007) as a pair of scan map observations using a 0.5$^\circ$ scan leg length and the medium scan rate (6.5~arcsec s$^{-1}$).  The raw data frames were reprocessed using the MIPS Data Analysis Tools version 3.10 \citep{gordon05} along with additional processing steps described by \citet{bendo12}.  The data processing steps applied to the individual data frames include a droop correction, dark current correction, electronic nonlinearity correction, detector responsivity correction, latent image removal, flatfielding, background subtraction, and asteroid removal.  An initial mosaic was created to identify and remove pixels with signals that were anomalously high or low in individual data frames compared to the rest of the dataset, and then a final mosaic was created for analysis.  We then subtracted any residual background from the image measured in locations outside the optical disc.  We do not apply any colour corrections to the final data, as the slope of the mid-infrared SED for NGC~4945 is not well-constrained and as the mid-infrared spectral slopes may vary significantly among starburst galaxies \citep{brandl06, casey12}.  The multiplicative colour corrections given by \citet{stansberry07} for power law spectra ranging from $\nu^{-3}$ and $\nu^3$ vary from 0.960 to 1.027, so we will treat this range of variation as an additional 3.5\% uncertainty in the 24~$\mu$m flux densities.

To check the overall flux calibration and data processing, we compared the global flux density to the flux density measured from the Infrared Astronomical Satellite (IRAS).  We measured the {\it Sptizer} 24~$\mu$m emission within a 24$\times$10~arcmin diameter region, which is large enough to encompass both the optical disc and emission related to extended structure in the beam that falls outside the optical disc.  The {\it Spitzer} 24~$\mu$m flux density is 31.9$\pm$1.6~Jy.  The globally-integrated IRAS 25~$\mu$m flux density reported by \citet{sanders03} is 42.3~Jy, which is $\sim$30\% higher than the {\it Spitzer} measurement.  While IRAS has a calibration uncertainty of 5\% at 25~$\mu$m \citep{wheelock94}, the uncertainties in flux density measurements for nearby galaxies are generally higher, and the flux densities measured by \citet{sanders03} may differ by up to 30\% compared to older measurements.  Additionally, while the colour corrections for the {\it Spitzer} data are small, the colour corrections for the IRAS data may range from $\sim$20\% to $\sim$70\% for power law spectra ranging from $\nu^{1}$ to $\nu^3$ \citep{wheelock94}, which would be typical of starburst galaxies.  The application of such a correction to the IRAS data would more closely match the IRAS and {\it Spitzer} measurements, although the correction is highly uncertain because we have not constrained the spectral slope.  In this context, the mismatch between the {\it Spitzer} and IRAS measurements does not seem to be a critical issue.

The final {\it Spitzer} 24~$\mu$m image appears free of major artefacts.  However, the centre of the galaxy appeared suppressed, either because of nonlinearity or saturation effects, within the central 3$\times$3 pixels in the individual raw data frames, which corresponds to a region with a radius of $\sim$3.8~arcsec in radius.  Even though the effects are removed early in the data processing, the measurements of the central flux density from the {\it Spitzer} 24~$\mu$m may be inaccurate, which is why we also use WISE 22~$\mu$m data in this analysis.  (The {\it Spitzer} 70 and 160~$\mu$m data are affected by more severe saturation effects, but the {\it Herschel} data at these wavelengths have no such problems.)

\subsubsection{WISE 22~$\mu$m data}

The WISE 22~$\mu$m data were acquired during the course of the WISE Cryogenic Survey described by \citet{wright10}.  The image that we use is the atlas tile 1974m500\_ac51 produced by the WISE Science Data System version 6.0 and released in the AllWISE Data Release.  This atlas tile is a standard 1$^\circ$.56 square.  NGC 4945 appears near the northwest edge of this image.  The southwest tip of the optical disc (a regions with a width of 1.5~arcmin) falls outside the tile, although this region contains $<$1\% of the emission seen in the {\it Spitzer} 24~$\mu$m image.  Since are primarily interested in the centre of the galaxy, which is $\sim$5.7~arcmin from the edge of the frame, the missing emission should not affect our results.  Before analysis, we subtracted a local background using measurements from regions on either side of the optical disc and outside the regions affected by the wings of the beam associated with the source.  The color corrections given by \citet{wright10} vary by $\ltsim$1\% from unity for power law spectra ranging from $\nu^{-3}$ to $\nu^3$, so we will not apply any colour corrections to the data.

We measured the globally-integrated WISE 22~$\mu$m flux density in the same 24$\times$10~arcmin diameter region used for the {\it Spitzer} 24~$\mu$m data (although this region partially falls outside the area covered in the atlas tile) and obtained a globally-integrated flux density of 35.8$\pm$0.7~Jy.  This falls within $\sim$10\% of the {\it Spitzer} 24~$\mu$m and IRAS 25~$\mu$m flux densities, which seems like a reasonable agreement.

\subsubsection{Herschel 70, 100, and 160~$\mu$m data}

\begin{figure*}
\begin{center}
\epsfig{file=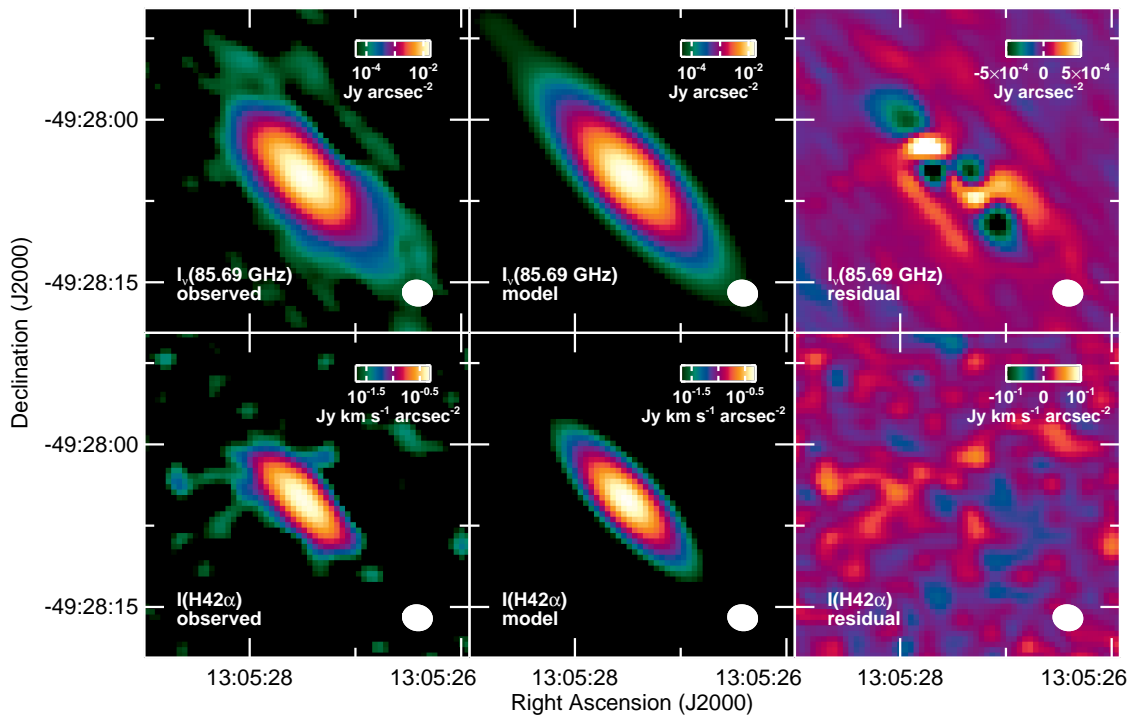}
\end{center}
\caption{Images of the 85.69~GHz continuum (top row) and the H42$\alpha$ line emission (bottom row) in the central 30~arcsec of NGC~4945.  The left column shows the observed images, the central column shows the best-fitting exponential functions, and the right column shows the residuals from the model fit.  Each image is oriented so that north is up and east is to the left.  The FWHM of the reconstructed beam (2.6$\times$2.3~arcsec, or 48$\times$42~pc for a distance of 3.8~Mpc) is shown in the lower right corner of each panel.  Logarithmic colour scales are used for the observed and model images, while linear scales are used for the residual images.}
\label{f_map_contline}
\end{figure*}

The 70, 100, and 160~$\mu$m image data were acquired with the Photoconductor Array Camera and Spectrometer \citep[PACS; ][]{poglitsch10} on {\it Herschel} in four observations.  Two observations (observations 1342203022 and 1342203023) were performed on 13 August 2010 and produced 100 and 160~$\mu$m images.  The other two observations (1342223660 and 1342223661) were performed on 04 July 2011 and produced 70 and 160~$\mu$m images.  Each pair of observations is a set of scan maps where the scans were performed at 20 arcsec s$^{-1}$ speeds in orthogonal directions to cover 30$\times$30~arcmin regions.

The individual PACS data frames were reprocessed using the standard PACS data pipelines within version 14.0.0 of the {\it Herschel} Interactive Processing Environment \citep[{\sc HIPE}; ][]{ott10}.  The pipeline data processing steps include masking bad pixels, saturated pixels, and pixels strongly affected by electrical crosstalk; converting the data units to Jy; and applying nonlinearity and instrument evaporator temperature corrections.  Images were created using {\sc JScanam}, the Java version of {\sc SCANAMORPHOS} \citep{roussel13}, which also removes variations in the background signal and cosmic rays from the data.  Residual offsets in the background emission were measured in regions outside the optical disc of the galaxy and then subtracted from the images.

As a check on the results, we measured flux densities (without colour corrections) within the 24$\times$10~arcmin diameter region used for the mid-infrared data and then compared the results to the globally-integrated flux densities from IRAS.  Unfortunately, the 100~$\mu$m band is the only waveband that is similar between the two instruments, so it is the only band where we can make a direct comparison.  The {\it Herschel} 100~$\mu$m flux density is $1880\pm60$~Jy, while the IRAS 100~$\mu$m flux density reported by \citet{sanders03} is 1330~Jy.  The difference is $\sim$30\%, which, as we mentioned in the comparison between {\it Spitzer} 24~$\mu$m and IRAS 25~$\mu$m data, is the same level of variation seen between the \citet{sanders03} and older measurements.  The {\it Herschel} 70~$\mu$m flux density is $1140\pm60$~Jy, which lies in between the IRAS 60~$\mu$m flux density of 625~Jy and the 100~$\mu$m flux density measurements.  The discrepancy between the {\it Herschel} and IRAS 100~$\mu$m is noted, but given the generally broad uncertainties in the IRAS data, we will not worry too much about the 30\% mismatch in the flux densities.

\subsubsection{Herschel 250, 350, and 500~$\mu$m data}

The 250, 350, and 500~$\mu$m data were acquired with the Spectral
and Photometric Imaging REceiver \citep[SPIRE; ][]{griffin10} instrument on {\it Herschel} in observation 1342203079, which was performed on 15 August 2010.  The observation is a large scan map of a 30$\times$15~arcmin region.  The region was scanned in two orthogonal directions at a 30 arcsec s$^{-1}$ speed with the detector voltages set to nominal mode.

The data were processed using the standard SPIRE pipeline in {\sc HIPE} version 14.0.0 and calibration tree spire\_cal\_14\_2.  The timeline data were processed through electrical crosstalk removal, cosmic ray removal, low pass filter, flux conversion, temperature drift removal, and bolometer time response correction steps; more details are given by \citet{dowell10}.  After this, the timeline data were processed through a destriper algorithm that removes offsets among the baseline signals in the individual bolometer timelines.  For display purposes, final images were created using the naive mapmaker in {\sc HIPE}.  Photometry is performed on the unresolved central source in the timeline data, so we applied no gain corrections for extended source emission.

\section{ALMA images}
\label{s_image}

Images of the 85.69~GHz continuum emission and integrated H42$\alpha$ emission within the central 30~arcsec of NGC~4945 are shown in the left column of Figure~\ref{f_map_contline}.  The emission in both of these images traces a relatively smooth, axisymmetric exponential disc, although the structure of the emission is somewhat noisier in the H42$\alpha$ image.  No additional emission is detected at above the 3$\sigma$ level outside the central 30~arcsec.  

To characterize the 85.69~GHz and H42$\alpha$ emission, we fit the images with a thin exponential disc that is projected at an inclined angle and convolved with a Gaussian function equivalent to the reconstructed beam using an updated version of the techniques from \citet{bendo06} and \citet{young09}.  Most other published results describe the spatial extent of the emission as a Gaussian function, which is not necessarily a physically accurate descriptions of the structure but which is easier to work with mathematically. For comparisons to these results, we also fit the data using an unprojected Gaussian function convolved with a second Gaussian function representing the reconstructed beam.

\begin{table*}
\centering
\begin{minipage}{100mm}
\caption{Parameters describing the disc models fit to the H42$\alpha$ and 85.69~GHz continuum data.}
\label{t_model}
\begin{tabular}{@{}llcc@{}}
\hline
Model &
    Quantity &
    85.69~GHz &
    H42$\alpha$\\
&
&
    Continuum &
    Line \\
\hline
Exponential Disc &
Major Axis Scale Length (arcsec) &
    $1.99 \pm 0.08$ &
    $2.33 \pm 0.09$ \\
&
Major Axis Scale Length (pc)$^a$ &
    $36 \pm 3$ &
    $42 \pm 4$\\
&
Minor/Major Axis Ratio &
    $0.23 \pm 0.03$ &
    $0.22 \pm 0.03$ \\
&
Major Axis Position Angle (deg)$^b$ &
    $48.0 \pm 2.0$ &
    $47.6 \pm 1.6$ \\
Gaussian Function &
Major Axis FWHM (arcsec) &
    $6.11 \pm 0.03$ &
    $7.30 \pm 0.09$ \\
&
Major Axis FWHM (pc)$^a$ &
    $112 \pm 9$ &
    $134 \pm 11$ \\
&
Minor Axis FWHM (arcsec) &
    $1.56 \pm 0.19$ &
    $1.80 \pm 0.06$ \\
&
Minor Axis FWHM (pc)$^a$ &
    $29 \pm 4$ &
    $33 \pm 3$ \\
&
Major Axis Position Angle (deg)$^b$ &
    $48.0 \pm 1.9$ &
    $47.5 \pm 0.5$ \\
\hline
\end{tabular}
$^a$ Dimensions in pc are calculated using a distance of 3.8~Mpc.  The uncertainties include the 0.3 Mpc uncertainty in the distance.\\
$^b$ The angles are degrees from north through east.
\end{minipage}
\end{table*}

The central column of Figure~\ref{f_map_contline} shows the model exponential discs fit to the data, and the right column shows the residual signal when the model is subtracted from the observed images.  Table~\ref{t_model} lists the parameters of the best-fitting exponential disc and Gaussian models.  Both the continuum and line images trace very similar structures, and the best-fitting parameters are also very similar.  Additionally, the central coordinates of the best fitting discs are within 0.2~arcsec of each other, which is a difference smaller than the beam or map pixel size.  The only significant difference is that the scale length of the line image is slightly larger than the scale length for the continuum image.  Both models replicate the observed pixel values to within 15\% where the H42$\alpha$ emission is detected at the 5$\sigma$ level.

The central disc observed in 85.69~GHz continuum emission and H42$\alpha$ line emission is similar to what has previously been observed in millimetre and radio data. The Gaussian FWHMs of the 85.69~GHz and H42$\alpha$ discs are $\sim$7$\times$1.7~arcsec, which are comparable to the deconvolved Gaussian FWHMs measured by \citet{ott01} at 1.4~GHz, by \citet{cunningham05} at 89~GHz, and by \citet{roy10} in H91$\alpha$ and H92$\alpha$ line emission.  The FWHM measured by \citet{chou07} is broader (9.8$\times$5.0~arcsec), but this could be because the band is dominated by emission from large dust grains, which could be more broadly distributed than the photoionized gas or synchrotron emission.  The HST images published by \citet{marconi00} show that the detected Pa$\alpha$ emission is much more asymmetric as a result of the heavy dust obscuration in the central starburst.  The Pa$\alpha$ emission also appears much more clumpy than the ALMA emission, although this is primarily a consequence of the higher resolution of the HST data.  

If photoionized gas around the central AGN was a significant line or continuum emission source, we would expect to see a central peak with emission significantly higher than what is predicted by the best-fitting exponential disc.  The absence of such a central peak implies that virtually all of the H42$\alpha$ and continuum emission originates from the circumnuclear starburst.  The AGN is potentially so heavily obscured that none of the photoionizing radiation escapes from the inner regions of the accretion disc.  This would be consistent with the analyses based on near- and mid-infrared data that found no evidence for ionization of the circumnuclear environment by a hard radiation field associated with an AGN \citep{marconi00, spoon00, spoon03, perezbeaupuits11}.

\section{ALMA spectra}
\label{s_spec}

\begin{figure}
\epsfig{file=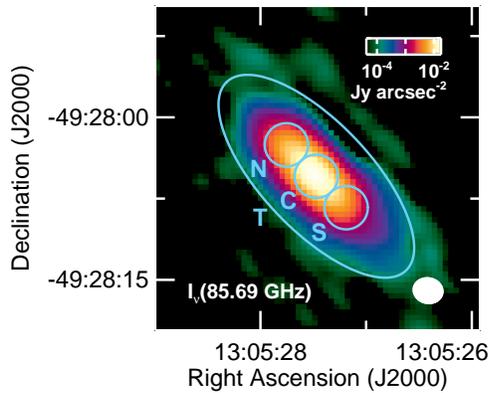,width=7.5cm}
\caption{The 85.69~GHz image marked with regions identifying the locations where spectra were measured.  The total (T) region is a 30$\times$12~arcsec diameter ellipse with a position angle of 43.5$^\circ$ that is aligned with the observed continuum emission.  The T region also encompasses all of the emission detected at $>$3$\sigma$ from the central region.  The north (N), centre (C), and south (S) regions have diameters of 4~arcsec.  See the caption for Figure~\ref{f_map_contline} for other details on the layout.}
\label{f_map_measreg}
\end{figure}

\begin{figure*}
\epsfig{file=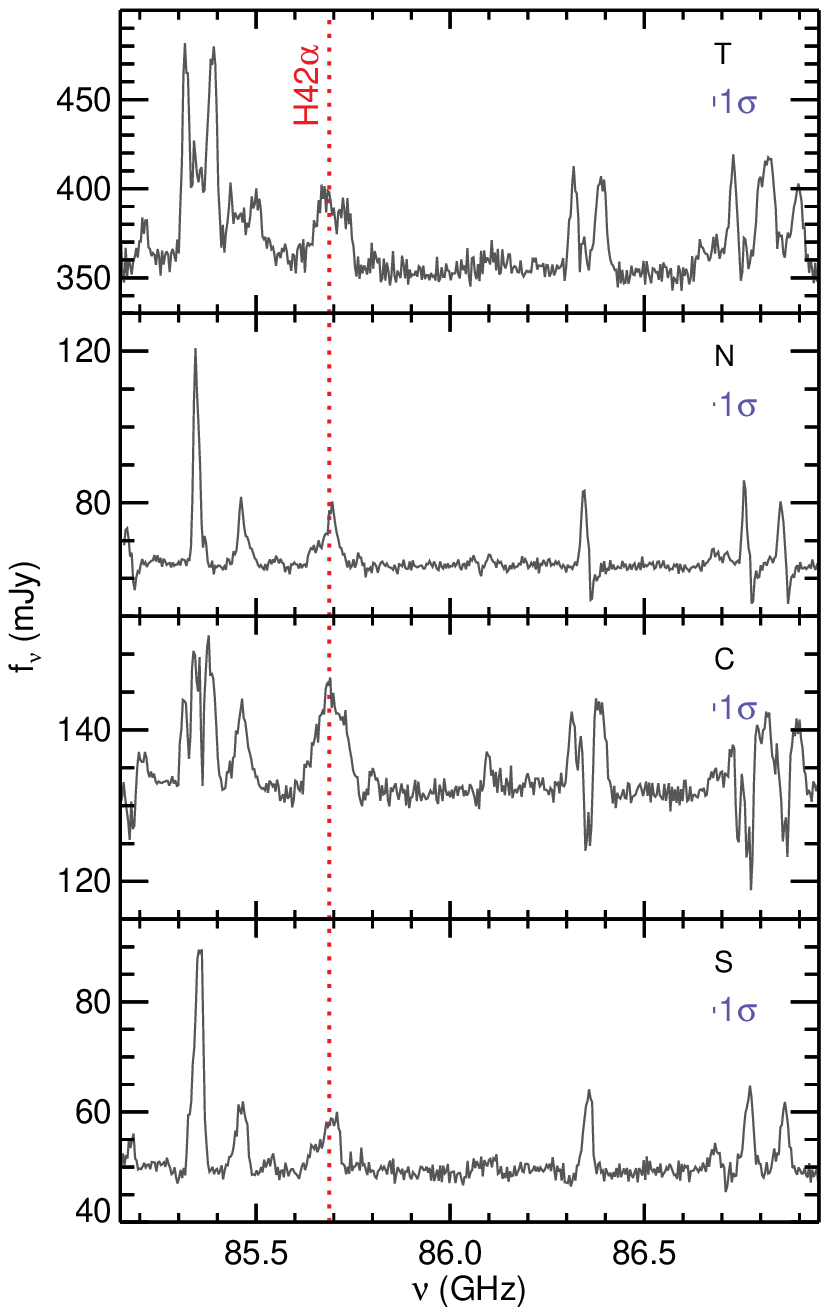}
~~~~
\epsfig{file=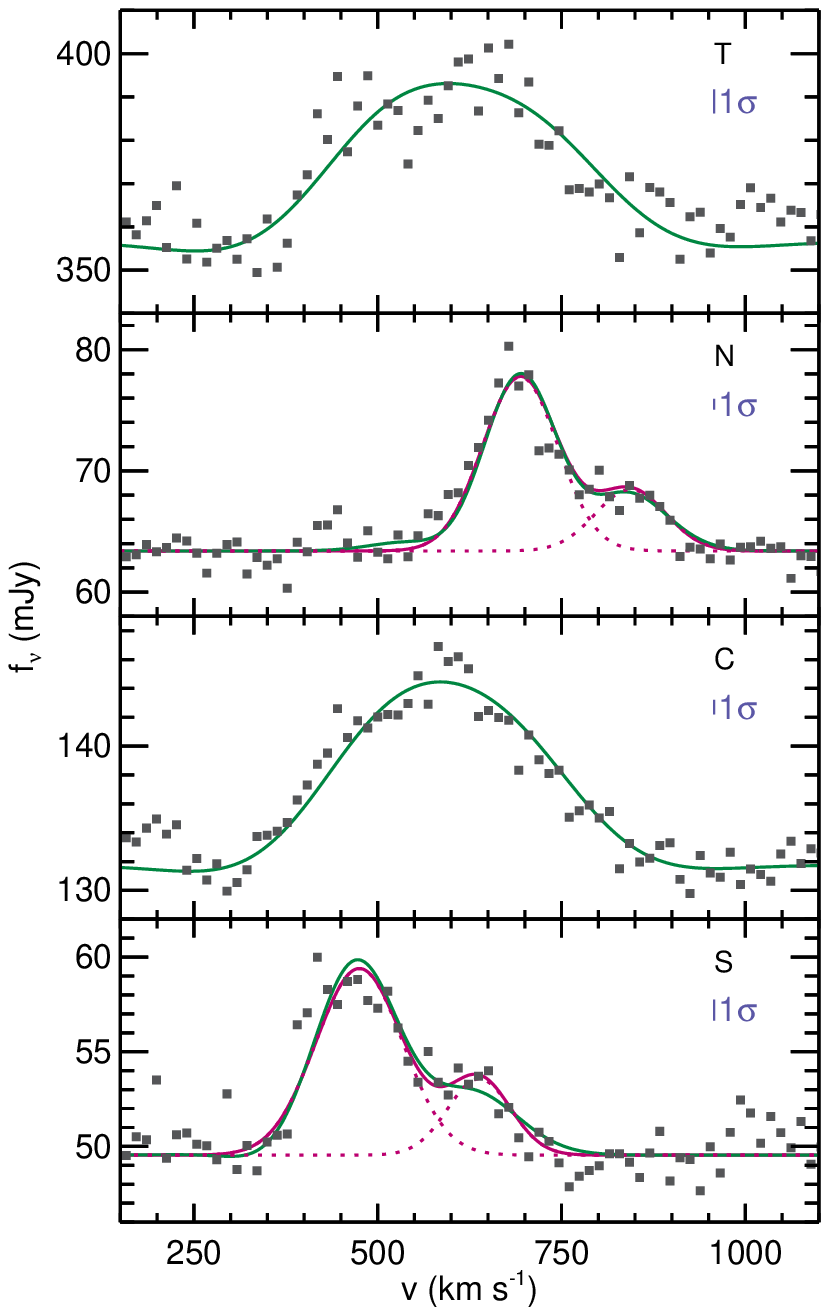}
\caption{Spectra for the regions shown in Figure~\ref{f_map_measreg}.  The panels on the left show the spectra measured within the entire spectral window.  The frequencies have been shifted to rest frequencies using the velocities listed in Table~\ref{t_gh}.  The position of the H42$\alpha$ line is marked with a red dotted line.  The spectra on the right show the data for the channels covering the H42$\alpha$ line.  The green lines show the best fitting Gauss-Hermite polynomials. The solid purple lines in the panels for the N and S regions show the profiles fit with two Gaussian functions with FWHM of 100-140~km~s$^{-1}$, and the dotted purple lines show the individual Gaussian components in these fits.  The parameters for all of these functions are listed in Table~\ref{t_gh}.  All velocities are in the barycentric frame and are calculated using the relativistic convention.  The blue bar on the upper right of each image shows the 1$\sigma$ uncertainties in the data.}
\label{f_spec}
\end{figure*}

Figure~\ref{f_map_measreg} shows locations where we extracted spectra in our data, and Figure~\ref{f_spec} shows both spectra for the entire spectral window and spectra for the channels near the H42$\alpha$ line.  Aside from the H42$\alpha$ line, multiple molecular lines can be seen in the spectra for the full spectral window.  The molecular line structure is quite complex, partly as a result of absorption of both line and continuum emission by molecular gas outside the inner region.  The molecular line emission is discussed in more detail by Henkel et al. (2016, in preparation).  The H42$\alpha$ emission does not show any of these effects, and it would be highly unusual for photoionized gas at these frequencies to be optically thick.

Although the H42$\alpha$ line emission is not optically thick, the line profiles are notably asymmetric, particularly in the north (N) and south (S) regions.  We will discuss these features below, but first we need to parameterize the non-Gaussian shapes of the lines.  While it is possible to fit the profiles with multiple Gaussian components, the results are not always reliable or repeatable, as multiple degenerate sets of parameters could be found for any given profile, as it is not always possible to identify the presence of a second Gaussian component, and as the results depend on our a priori assumptions.  While it is possible to adjust the fitting to obtain reliable results when fitting two Gaussian funtions to the asymmetric line profiles in Figure~\ref{f_spec}, it is impractical to do this for the individual pixels in the image cube.

Given this, we fit both the image cube data and the spectra in Figure~\ref{f_spec} using Gauss-Hermite functions as described by \citet{vandermarel93}.  The Gauss-Hermite functions are modified Gaussian functions that include additional dimensionless terms for skewness ($h_3$) and kurtosis ($h_4$).   The functions including these skewness and kurtosis terms can be described by
\begin{equation}
I_\nu(v)=A \frac{\alpha(w)}{\sigma} [1 + h_3 H_3(w) + h_4 H_4(w)],
\end{equation}
where
\begin{equation}
w = (v - v_0) / \sigma,
\end{equation}
\begin{equation}
\alpha(w) = \frac{1}{\sqrt{2 \pi}} e^{-w^2/2},
\end{equation}
\begin{equation}
H_3(w) = \frac{1}{\sqrt{6}}(2 \sqrt{2} w^3 - 3 \sqrt{2} w),
\end{equation}
\begin{equation}
H_4(w) = \frac{1}{\sqrt{24}}(4 w^4 - 12 w^2 + 3).
\end{equation}
A positive $h_3$ value indicates that the wing of the Gauss-Hermite function at values above $v_0$ is much broader than the wing for values below $v_0$, while a negative value indicates the reverse.  A positive $h_4$ value indicates that the peak is very narrow and both wings are very broad compared to a Gaussian function, while a negative $h_4$ value indicates that the peak is very broad and the wings are truncated.

We fitted these functions to all continuum-subtracted data between channels 310 and 365 (sky frequencies 85.401-85.616~GHz; barycentric velocities 254-1007 km s$^{-1}$) in the continuum-subtracted image cubes.  The amplitude ($A$), mean velocity ($v_0$), Gaussian width parameter ($\sigma$), $h_3$, and $h_4$ are all treated as free parameters.  We also report FWHM for the data, but since the FWHM of the line emission does not scale linearly with $\sigma$ in Gauss-Hermite functions, the FWHMs are determined by directly measuring of the widths of the best fitting functions.  

\begin{table*}
\centering
\begin{minipage}{143mm}
\caption{Parameters for functions fit to the H42$\alpha$ lines.}
\label{t_gh}
\begin{tabular}{@{}lcccccccc@{}}
\hline
Region &
    \multicolumn{4}{c}{\multirow{2}{*}{Gauss-Hermite Functions}} &
    \multicolumn{4}{c}{Two Gaussian Functions}\\
&
    &
    &
    &
    &
    \multicolumn{2}{c}{Lower $v_0$ Component} &
    \multicolumn{2}{c}{Higher $v_0$ Component} \\
&
    $v_0^a$ &
    $\sigma$ &
    $h_3$ &
    $h_4$ &
    $v_0^a$ &
    $FWHM$ &
    $v_0^a$ &
    $FWHM$ \\
&
    (km s$^{-1}$) &
    (km s$^{-1}$) &
    &
    &
    (km s$^{-1}$) &
    (km s$^{-1}$) &
    (km s$^{-1}$) &
    (km s$^{-1}$) \\
\hline
Total (T) &
    $611 \pm 7$ &
    $133 \pm  6$ &
    $0.03 \pm 0.04$ &
    $-0.10 \pm 0.04$ &
    &
    &
    &
   \\
North (N) &
    $706 \pm 2$ &
    $ 69 \pm 2$ &
    $0.19 \pm 0.02$ &
    $0.17 \pm 0.02$ &
    $694 \pm 2$ &
    $119 \pm 7$ &
    $844 \pm  9$ &
    $112 \pm 14$ \\
Centre (C) &
    $592 \pm 3$ &
    $124 \pm 2$ &
    $0.02 \pm 0.02$ &
    $-0.07 \pm 0.02$ &
    &
    &
    &
   \\
South (S) &
    $489 \pm 4$ &
    $ 77 \pm 3$ &
    $0.21 \pm 0.04$ &
    $0.10 \pm 0.05$ &
    $475 \pm 13$ &
    $138 \pm 26$ &
    $638 \pm 14$ &
    $102 \pm 29$ \\
\hline
\end{tabular}
$^a$ The mean velocities are relativistic velocities in the barycentric frame.
\end{minipage}
\end{table*}

\begin{figure*}
\epsfig{file=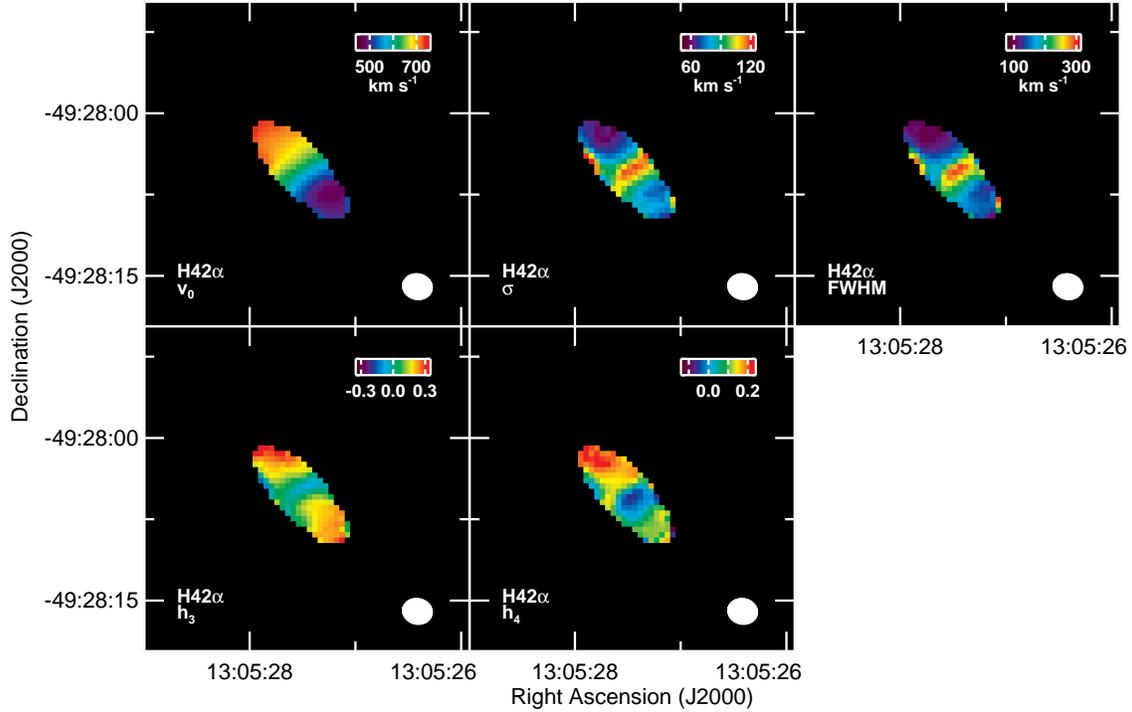}
\caption{Images of the mean velocity ($v_0$), Gaussian width parameter ($\sigma$), FWHM, $h_3$, and $h_4$ for the H42$\alpha$ line in each image cube pixel where the integrated H42$\alpha$ flux is detected at the 5$\sigma$ level.  The velocities are in the barycentric frame.  See the caption for Figure~\ref{f_map_contline} for other details on the layout.}
\label{f_map_gh}
\end{figure*}

Table~\ref{t_gh} lists the parameters describing the Gauss-Hermite function fit to the profiles shown in Figure~\ref{f_spec}.  Additionally, Figure~\ref{f_map_gh} shows maps of these parameters for all data where the integrated H42$\alpha$ line emission is detected at above the $5\sigma$ level. The velocities are consistent with the rotation of a disc around a central source, and the increased dispersion in the centre is expected.  The FWHM of the central region is $<$300 km s$^{-1}$, much less than would be expected for high-velocity photoionized gas associated with type 1 AGN, as is discussed by \citet{izumi16}.  

The mean barycentric velocity of 611$\pm$7~km~s$^{-1}$ based on the total integrated H42$\alpha$ emission for the disc is consistent with the velocity of 581$\pm$15~km~s$^{-1}$ based on the H91$\alpha$ and H92$\alpha$ lines from \citet{roy10}. The H42$\alpha$ mean disc velocity is also within $\sim$3\% of the velocity of $\sim$590~km~s$^{-1}$ (converted from the local standard of rest (LSR) frame to the barycentric frame using $v_{bary}$=$v_{LSR}$+4.6~km~s$^{-1}$) based on multiple CO (2-1) lines observed from the central region as reported by \citet{chou07}.  However, velocities based on globally-averaged values, including velocities based on H$\alpha$ lines \citep[e.g.][]{mathewson96} or H{\small I} lines \citep[e.g.][]{mathewson96, ott01}, tend to be lower by $\sim$50~km~s$^{-1}$.  This is a known peculiarity in the velocity measurements for this galaxy, as also noted by \citet{chou07} and \citet{roy10}.

The most peculiar features in the line profiles are the skewed features to the east and west of the central region.   Skewing in spectral line profiles could be associated with gas or stars within the disc with different dynamical characteristics from the rest of the gas and stars (related to, for example, accreted gas and stars) that are rotating at a different velocity or even counterrotating relative to the rest of the disc.  To illustrate how the skewed line profiles could potentially be divided into separate velocity components, we fit the N and S regions with two Gaussian functions with widths of 100-140~mk~s$^{-1}$, as seen in Figure~\ref{f_spec}.  The two velocity components are separated by $\sim$150~km~s$^{-1}$.  (The line profiles of the C and T regions cannot be straightforwardly separated into two velocity components, which is why we do not fit the data with two Gaussian functions.)

In situations where skewed spectral line profiles are observed within galaxies, the profiles are expected to look like mirror images of each other in the east and west sides of the disc, with $h_3$ appearing positive on one side of the centre and negative on the other \citep[e.g.][]{vandermarel93, fisher97, bendo00, milosavljevic01, naab06, jesseit07}.  In NGC~4945, however, the $h_3$ is positive for regions on both the east and west side of the centre, which means that higher velocity gas is seen on both sides of the disc.  This makes it unlikely that the skewing is caused by accreted gas or stars.  Photoionization $\alpha$ lines from helium and carbon are also found near the H42$\alpha$ line and could make the lines appear skewed, but these lines would appear at higher frequencies and lower velocities, which is opposite of where we observe the broadening in the emission line.  Spectral lines listed by {\sc Splatalogue}\footnote{Accessible at http://www.cv.nrao.edu/php/splat/ .} at frequencies up to 0.05~GHz lower than the H42$\alpha$ line originate mainly from either very complex organic molecules, species with very high excitation states, or molecules with rare isotopes, all of which have no counterparts elsewhere in the spectrum and are otherwise unlikely to be seen in these star forming regions. The most likely remaining possibility is that the skewing is related to additional photoionized gas outside of the disc but projected along the line of sight.  The regions with high $h_3$ values in Figure~\ref{f_map_gh} correspond to regions with positive values in the residual maps in Figure~\ref{f_map_contline} where the exponential disc models have been subtracted from the observed data, which would be expected if the emission originated from clouds outside the disc.

\section{Spectral energy distribution}
\label{s_sed}

Even though free-free emission is generally the dominant emission source at 85.69~GHz \citep[e.g.][]{peel11}, synchrotron and thermal dust emission could contribute a significant fraction of the emission at this frequency \citep[e.g.][]{bendo15b}.  To use the continuum emission to estimate both $T_e$ and SFR, we need to remove the other continuum emission sources from the free-free emission.  This is best done by fitting the SED with multiple power laws representing the synchrotron, free-free, and dust emission.  

For this task, we will use all continuum data between 3 and 350~GHz for any measurement region comparable to the central source we detect at 85.69~GHz (where the emission at $>$3$\sigma$ falls entirely within a 30$\times$12~arcsec region).  The data used in the SED analysis are listed in Table~\ref{t_sed}.  We did not use data at $<$3~GHz because the slope of the synchrotron emission flattens, probably because the cosmic rays producing the higher frequency emission are more strongly affected by aging effects \citep[e.g.][]{davies06}.  We did not use data at $>$350~GHz because the data up to 350~GHz are sufficient for constraining the contribution of dust emission to the SED and because data at higher frequencies may deviate from a Rayleigh-Jeans function.  As implied above, we did not use flux density measurements integrated over the entire galaxy or over regions $>$1~arcmin.  We also excluded the data from the Australia Telescope 20 GHz Survey \citep{murphy10} because the flux densities are systematically low compared to all other measurements at similar frequencies, possibly because the algorithms used by the survey for flux density measurements were optimized for unresolved sources smaller than the central disc seen in the ALMA data.  \citet{lenc09} list multiple measurements at 17-23~GHz that we have averaged into a single number at 20~GHz.  We also include the 85.69~GHz flux density that we measure within the T aperture in Figure~\ref{f_map_measreg}, which is $0.36 \pm 0.02$~Jy.

\begin{table}
\caption{Data used for SED analysis.}
\label{t_sed}
\begin{center}
\begin{tabular}{@{}ccc@{}}
\hline
Frequency &
  Flux Density &
  Reference \\
(GHz) &
  (Jy) &
  \\
\hline
4.8 &
  3.055 &
  \citet{healey07}\\
4.8 &
  $2.7\pm0.3$ &
  \citet{lenc09}\\
5.0 &
  $2.370 \pm 0.001$ &
  \citet{forbes98}\\
8.33 &
  $1.4 \pm 0.2$ &
  \citet{roy10}\\
8.4 &
  1.0802 &
  \citet{healey07}\\
8.6 &
  $1.30\pm0.13$ &
  \citet{lenc09}\\
20 &
  $0.73\pm0.04$ &
  \citet{massardi08}\\
20 &
  $0.84\pm0.08^a$ &
  \citet{lenc09}\\
85.69 &
  $0.36 \pm 0.02$ &
  [this work] \\
230 &
  $1.3 \pm 0.3$ &
  \citet{wang04}\\
230 &
  $1.3 \pm 0.2$ &
  \citet{chou07}\\
345 &
  $7.2\pm0.8$ &
  \citet{weiss08}\\
\hline
\end{tabular}
\end{center}
$^a$ The number here represents an average of the 17-23~GHz flux density measurements published by \citet{lenc09}.\\ 
\end{table}

We model the synchrotron, free-free, and thermal dust emission components as power laws.  The spectral index of the synchrotron emission is treated as a free parameter.  The free-free emission is set to scale with the function for the Gaunt factor given by \citet{draine11}, which is
\begin{equation}
  g_{ff} = 0.5535 \ln \left| 
  \left[ \frac{T_e}{\mbox{K}} \right]^{1.5}
  \left[ \frac{\nu}{\mbox{GHz}} \right]^{-1}
  Z^{-1}
  \right| -1.682.  
\end{equation}  
In this equation, $Z$ is the charge of the ions, which we will assume to be $\sim$1.  For fitting the SED, we set $T_e$ to 5000~K for fitting the SED, although we discuss below what happens when other values of $T_e$ are used.  The spectral index of the dust emission is fixed to 4, which is what is expected if the dust has an emissivity that is proportional to $\nu^{2}$ as is commonly assumed in many dust emission models \citep[e.g.][]{draine03}.  Analyses using {\it Herschel} data had suggested that dust emissivity may be proportional to $\nu^{1.5}$ \citep{boselli12}, but the slope of the 230 and 345~GHz data in Table~\ref{t_sed} is more consistent with dust emissivities that are proportional to $\nu^{2}$.  

\begin{figure}
\epsfig{file=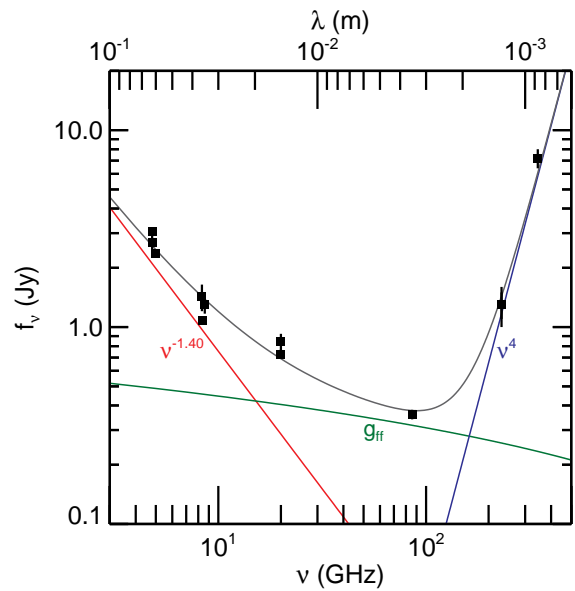}
\caption{The SED for the central disc in NGC~4945.  The data plotted here are listed in Table~\ref{t_sed}.  The black line shows the best fitting SED model for the data.  The red line represents synchrotron emission with a spectral index of -1.40; this index was a free parameter in the fit.  The green line represents free-free emission with a spectral index fixed at -0.118.  The blue line corresponds to thermal dust emission with a spectral index fixed at 4.}
\label{f_sed}
\end{figure}

Figure~\ref{f_sed} shows the power laws that we fit to the data listed in Table~\ref{t_sed}.  According to the fits, the free-free emission is the dominant emission source between 20 and 150~GHz.  At 85.69~GHz, the free-free emission produces 84\% of the total observed emission, synchrotron emission produces 10\% of the total, and thermal dust emission produces 6\%.  Using a Monte Carlo approach\footnote{For the Monte Carlo analysis, we used 10\% uncertainties for the \citet{forbes98} data, where the uncertainties seem unrealistically low and appear to only contain instrumental uncertainties, and for the \citet{healey07} data, where no uncertainties are reported.}, we found that the uncertainty in the percent of total emission from free-free emission is 7\%.  We performed several tests using $T_e$ ranging from 3000 to 10000~K for the free-free emission and spectral indices ranging from 3.5 to 4 for the dust emission, and we also performed tests where we represented the free-free emission with power laws with spectral indices ranging from -0.1 to -0.2.  In all of these tests, we found that the percent of the flux density from free-free emission at 85.69~GHz varied by $\leq$10\%.

Given these results, we will proceed with calculations of $T_e$ and SFR based on applying a multiplicative correction of 0.84$\pm$0.10 to the 85.69~GHz continuum emission.  However, we will also discuss the values of $T_e$ and SFR that we would derive if we applied no correction and if we applied a correction of 0.50.

\section{Electron temperature}
\label{s_te}

The hydrogen recombination line emission $f_\nu$(line) integrated over velocity $v$ can be calculated using
\begin{equation}
\begin{split}
  \frac{\int f_\nu(\mbox{line}) dv}{\mbox{Jy km s}^{-1}}
  = 2.50\times10^{-31} 
  \ \ \ \ \ \ \ \ \ \ \ \ \ \ \ \ \ \ \ \ \ \ \ \ \ \ \ \ \ \ \ \ \ \ \ \ \ \ \ \ \ \ \ \ \ \ \ \\
  \times n_e n_p V 
  \left[\frac{\epsilon_\nu}{\mbox{erg s}^{-1}\mbox{ cm}^{-3}}\right]
  \left[\frac{\nu}{\mbox{~GHz}}\right]^{-1}
  \left[\frac{D}{\mbox{ Mpc}}\right]^{-2}
\end{split}
\label{e_fh42a}
\end{equation}
from \citet{scoville13}, where $n_e$ and $n_p$ are the electron and proton densities, $V$ is the volume, $\epsilon_\nu$ is the emissivity from \citet{storey95}, $\nu$ is the frequency, and $D$ is the distance.   The $\epsilon_\nu$ term does not vary significantly as a function of electron density between $10^2$ and $10^5$ cm$^{-1}$ but varies strongly as a function of $T_e$ in the 1000-10000~K range.  

The free-free emission $f_\nu$(cont) can be determined using
\begin{equation}
\begin{split}
  \frac{f_\nu(\mbox{cont})}{\mbox{Jy}} =     
  \ \ \ \ \ \ \ \ \ \ \ \ \ \ \ \ \ \ \ \ \ \ \ \ \ \ \ \ \ \ \ \ \ \ \ \ \ \ \ \ \ \ \ \ \ \ \ \ \ \ \ \ \ \ \ \ \ \ \ \ \ \ \ \ \ \ \ \ \ \ \ \ \ \ \ \ \ \ \\
  5.70\times10^{-65} g_{ff} Z^{2} n_e n_p V
  \left[\frac{T_e}{\mbox{~K}}\right]^{-0.5}
  \left[\frac{D}{\mbox{ Mpc}}\right]^{-2},
\end{split}
\label{e_fdcont}
\end{equation}
which is derived for $hv \ll kT$ from the equations given by \citet{draine11}.

\begin{table*}
\centering
\begin{minipage}{95mm}
\caption{$T_e$ analysis results.}
\label{t_te}
\begin{tabular}{@{}ccccc@{}}
\hline
Region &
  85.69~GHz Free-Free &
  H42$\alpha$ &
  Line / Free-Free &
  $T_e$\\
&
  Flux Density &
  Flux &
  Emission Ratio &
  (K)\\
& 
  (Jy) &
  (Jy km s$^{-1}$) &
  (Jy km s$^{-1}$ / Jy) &
  \\
\hline
T &
  $0.31 \pm 0.04$ &
  $12.9 \pm 0.8$ &
  $42 \pm 5$ &
  $5400 \pm 600$ \\
N &
  $0.054 \pm 0.007$ &
  $2.5 \pm 0.1$ &
  $46 \pm 6$ &
  $5000 \pm 600$ \\
C &
  $0.111 \pm 0.015$ &
  $3.9 \pm 0.2$ &
  $35 \pm 4$ &
  $6300 \pm 700$ \\
S &
  $0.042 \pm 0.006$ &
  $1.9 \pm 0.1$ &
  $45 \pm 5$ &
  $5100 \pm 600$ \\
\hline
\end{tabular}
$^a$ The free-free flux densities are based on the continuum flux densities in these apertures multiplied by $0.84\pm0.10$.
\end{minipage}
\end{table*}

\begin{figure}
\epsfig{file=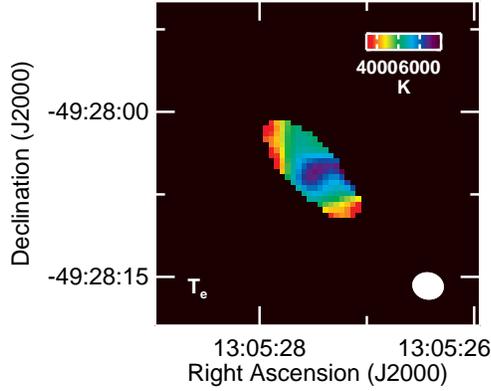,width=7.5cm}
\caption{Map of $T_e$ where the integrated H42$\alpha$ flux is detected at the 5$\sigma$ level.  See the caption for Figure~\ref{f_map_contline} for other details on the layout.}
\label{f_map_te}
\end{figure}

Using these two equations, we can write the ratio of the line-to-continuum emission as
\begin{equation}
\begin{split}
  \frac{\int f_\nu(\mbox{line}) dv}{f_\nu(\mbox{cont})} 
  \left[\frac{\mbox{Jy}}{\mbox{Jy km s}^{-1}}\right] = 
  \ \ \ \ \ \ \ \ \ \ \ \ \ \ \ \ \ \ \ \ \ \ \ \ \ \ \ \ \ \ \ \ \ \ \ \ \ \ \ \ \ \ \ \ \ \ \ \ \\
  4.38\times10^{33} g_{ff}^{-1}
  \left[\frac{\epsilon_\nu}{\mbox{erg s}^{-1}\mbox{ cm}^{-3}}\right]
  \left[\frac{\nu}{\mbox{~GHz}}\right]^{-1}
  \left[\frac{T_e}{\mbox{~K}}\right]^{0.5},
\label{e_te}
\end{split}
\end{equation}
which we use to derive $T_e$.  When doing so, we use $\epsilon_\nu$ values corresponding to case B recombination and a fixed $n_e$ value of $10^3$ cm$^{-3}$ but interpolate between values of $\epsilon_\nu$ for different $T_e$.  The value of $\epsilon_\nu$ varies inversely with $T_e$, so the ratio given in Equation~\ref{e_te} also decreases with $T_e$ in spite of the $T_e$ term.  The $g_{ff}$ term is also allowed to vary with $T_e$ for this analysis, but the variations are only $\sim$1.4$\times$ over the range of 3000-10000~K.  All $T_e$ are derived assuming that 84$\pm$10\% of the continuum emission originates from free-free emission.

Table~\ref{t_te} lists electron temperatures measured in the different regions shown in Figure~\ref{f_map_measreg}, and Figure~\ref{f_map_te} shows a map of $T_e$.  We obtain a $T_e$ of 5400$\pm$600~K for the emission integrated over the entire disc.  In the central region, $T_e$ increases slightly to $6300\pm700$~K, while in the rest of the disc, $T_e$ is $\sim$5000~K.  This increase of 1300~K is akin to the increase of 600-800~K in $T_e$ seen in the centre of NGC~253 \citep{bendo15b}, which is an object with no detected AGN.  None the less, we examined whether the increase in $T_e$ in the central region of NGC~4945 could still be the result of increased synchrotron emission from the central AGN.  If we assume that the C region should have the same line/continuum ratio as the outer disc when no excess synchrotron emission is present, then we estimate that $\sim$30\% of the continuum of the central source is synchrotron emission.   Alternately, if we assume that all of the 85.69~GHz synchrotron emission as estimated in the SED analysis for the T region in Section~\ref{s_sed} actually originates from the very centre of the galaxy, then $T_e$ would be 5500~K.  However, as described in Section~\ref{s_image}, the Gaussian FWHMs that we measure at 85.69~GHz are very similar to what has been measured at lower frequencies that are dominated by synchrotron emission, which would imply that the synchrotron emission should not appear more centrally concentrated than the free-free emission.  It is more likely that the changes in the observed line/continuum ratio reflect actual changes in $T_e$.

The $T_e$ measurements are broadly consistent with the low end of $T_e$ measurements within the Milky Way given by \citet{shaver83} and \citet{paladini04}.  If we applied no correction for 85.69~GHz emission originating from dust or synchrotron emission, we would measure a globally-integrated $T_e$ value of 6300~K, which would fall within the midrange of Milky Way $T_e$ measurements.  If only 50\% of the emission originated from free-free emission, then the globally-integrated $T_e$ would be 3300~K, which is very low compared to Milky Way measurements and also seems nearly unrealistically low for photoionized gas.  It would therefore seems unlikely that $\leq$50\% of the 85.69~GHz emission is produced by free-free emission.

\section{Star formation rates}
\label{s_sfr}

\subsection{Star formation rates from ALMA data}
\label{s_sfr_alma}

Equations \ref{e_fh42a} and \ref{e_fdcont} can be used to relate the observed fluxes to a photoionizing photon production rate $Q$ by applying
\begin{equation}
  Q=\alpha_B n_e n_p V
\label{e_emtoq}
\end{equation}
from \citet{scoville13}, where $\alpha_B$ is the effective recombination coefficient listed by \citet{storey95}.  The $\alpha_B$ term depends on $T_e$, varying by a factor of $\sim$4 between 3000 and 15000~K, but it does not vary significantly with $n_e$ for densities ranging from $10^2$ to $10^5$~cm$^{-3}$.  For our analysis, we will use $\alpha_B$ corresponding to the $T_e$ derived in Section~\ref{s_te} and $n_e$=$10^3$~cm$^{-3}$.

To convert $Q$ to SFR, we use
\begin{equation}
  \frac{\mbox{SFR}}{\mbox{M}_\odot~\mbox{yr}^{-1}}
   =5.41\times10^{-54}\frac{Q}{\mbox{s}^{-1}},
\label{e_qtosfr}
\end{equation}
which was derived by \citet{bendo15b} using version 7.0.1 of the {\sc Starburst99} models \citep{leitherer99, leitherer14}.  The conversion was derived using $Z=0.040$ metallicity\footnote{In this metallicity system, solar metallicity is $Z=0.020$.}, a \citet{kroupa02} initial mass function (IMF) for a mass range of 0.1-100 M$_\odot$, and the average of results from the Geneva evolutionary tracks that include no stellar rotation and rotation at 40\% of the break-up velocity\footnote{As discussed by \citet{leitherer14}, stellar rotation will enhance convection, which makes the stars hotter and more luminous.  The actual stellar rotation velocities are expected to be between the two extremes in the  {\sc Starburst99} models.}  The SFR is assumed to be continuous and to have been ongoing for greater that 3~Myr, which would be consistent with the presence of both supernovae (as detected by \citet{lenc09}) and photoionizing stars.

For comparison, the coefficient in the conversion given by \citet{kennicutt98} is a factor of $\sim$2 higher and the coefficient given by \citet{murphy11} \citep[see also ][]{calzetti07} is $\sim$35\% higher.  The difference between the \citet{murphy11} and \citet{bendo15b} coefficients is mainly a result of the use of models with rotating stars by \citet{bendo15b}.  If no rotation was included in the derivation of the relation between SFR and $Q$, the coefficient in Equation~\ref{e_qtosfr} would be very similar to the \citet{murphy11} coefficient.  The additional difference between the \citet{kennicutt98} and \citet{bendo15b} coefficients is related to the difference between the \citet{kennicutt98} and \citet{murphy11} coefficients.  As discussed by \citet{calzetti07}, this could be attributed to assumptions regarding the IMF and star formation history.  These differences are important to describe here because many of the conversions between mid-infrared flux and SFR are based on the \citet{kennicutt98} relation, and this could create a discrepancy between the measurements from the ALMA data and the mid-infrared data.

Using the relations in Equations~\ref{e_emtoq} and \ref{e_qtosfr}, we can rewrite Equations~\ref{e_fh42a} and \ref{e_fdcont} to give the following relations between observed line or continuum emission and star formation:
\begin{equation}
\begin{split}
  \frac{\mbox{SFR}(\mbox{line})}{\mbox{M}_\odot\mbox{ yr}^{-1}}=
  2.16\times10^{-23}
  \left[\frac{\alpha_B}{\mbox{ cm}^3\mbox{ s}^{-1}}\right]
  \ \ \ \ \ \ \ \ \ \ \ \ \ \ \ \ \ \ \ \ \ \ \ \ \ \ \ \ \ \ \\
  \times
  \left[\frac{\epsilon_\nu}{\mbox{erg s}^{-1}\mbox{ cm}^{-3}}\right]^{-1}
  \left[\frac{\nu}{\mbox{~GHz}}\right]
  \left[\frac{D}{\mbox{ Mpc}}\right]^{2}
  \left[\frac{\int f_\nu(\mbox{line}) dv}{\mbox{Jy km s}^{-1}}\right]
\end{split}
\end{equation}
\begin{equation}
\begin{split}
\frac{\mbox{SFR}(\mbox{cont})}{\mbox{M}_\odot\mbox{ yr}^{-1}}=
  9.49\times10^{10}
  \ \ \ \ \ \ \ \ \ \ \ \ \ \ \ \ \ \ \ \ \ \ \ \ \ \ \ \ \ \ \ \ \ \ \ \ \ \ \ \ \ \ \ \ \ \ \ \ \ \ \ \ \\
  \times g_{ff}^{-1}
  \left[\frac{\alpha_B}{\mbox{ cm}^3\mbox{ s}^{-1}}\right]
  \left[\frac{T_e}{\mbox{~K}}\right]^{0.5}
  \left[\frac{D}{\mbox{ Mpc}}\right]^{2}
  \left[\frac{f_\nu(\mbox{cont})}{\mbox{Jy}}\right] .
\end{split}
\label{e_sfr_freefree}
\end{equation}
We calculated SFRs for all regions in Figure~\ref{f_map_measreg} using the $T_e$ derived in Section~\ref{s_te} and assuming that 84$\pm$10\% of the continuum emission is from free-free emission.  Because the ratio of the line and continuum emission are coupled together in the calculation of $T_e$, the SFR measurements from the line and continuum are effectively coupled together.

\begin{table}
\caption{SFR based on ALMA data.}
\label{t_sfr_alma}
\begin{center}
\begin{tabular}{@{}ccc@{}}
\hline
Region &
  \multicolumn{2}{c}{SFR} \\
&
  85.69~GHz free-free &
  H42$\alpha$ line\\
&
  (M$_\odot$ yr$^{-1}$) &
  (M$_\odot$ yr$^{-1}$)\\
\hline
T &
  $4.42 \pm 0.49$ &
  $4.29 \pm 0.07$ \\
N &
  $0.80 \pm 0.22$ &
  $0.80 \pm 0.07$ \\
C &
  $1.48 \pm 0.44$ &
  $1.39 \pm 0.09$ \\
S &
  $0.62 \pm 0.15$ &
  $0.62 \pm 0.07$ \\
\hline
\end{tabular}
\end{center}
\end{table}

The SFRs are listed in Table~\ref{t_sfr_alma}.  The average of the continuum and line SFRs integrated over the central disc (the T region) and the associated uncertainty is 4.35$\pm$0.25~M$_\odot$ yr$^{-1}$; we will use this number in comparison to SFRs from the infrared and radio data.  Although the emission peaks in the centre of the disc, only about one-third of the total SFR is within the C region.  The SFRs are moderately dependent upon the fraction of continuum emission from free-free emission determined in Section~\ref{s_sed}.  If the continuum emission was purely from free-free emission, then the SFRs would increase by $\sim$10\%,  If half the continuum emission originated from free-free emission (which, as we discussed in Section~\ref{s_te}, leads to questionably low $T_e$), then SFRs would decrease by $\sim$20\%.

\subsection{Star formation rates from mid-infrared flux densities}
\label{s_sfr_mir}

We used the {\it Spitzer} 24~$\mu$m and WISE 22~$\mu$m images to derive mid-infrared based SFRs for the central disc.  As described in Section~\ref{s_data_ir}, the angular resolution of the {\it Spitzer} data is superior to the WISE data, but flux densities measured in the {\it Spitzer} data may be affected by nonlinearity or saturation effects even though such artefacts are hardly visible in the final image.  The central region is marginally resolved in the {\it Spitzer} image and is completely unresolved in the WISE image, so we treat the source as a point source when measuring flux densities; the resulting SFRs from this source should be comparable to what is measured in the T region in Figure~\ref{f_map_measreg}.  As the ellipticity of the central starburst is hardly discernible in these data, we use circular apertures for the photometry.  Infrared emission is also seen in compact star forming regions as well as diffuse structures distributed throughout the optical disc, all of which is at least 50$\times$ lower in surface brightness than the peak surface brightness of the central source.  To deal with this extended emission, we experimented with applying local background subtractions.

\begin{figure}
\epsfig{file=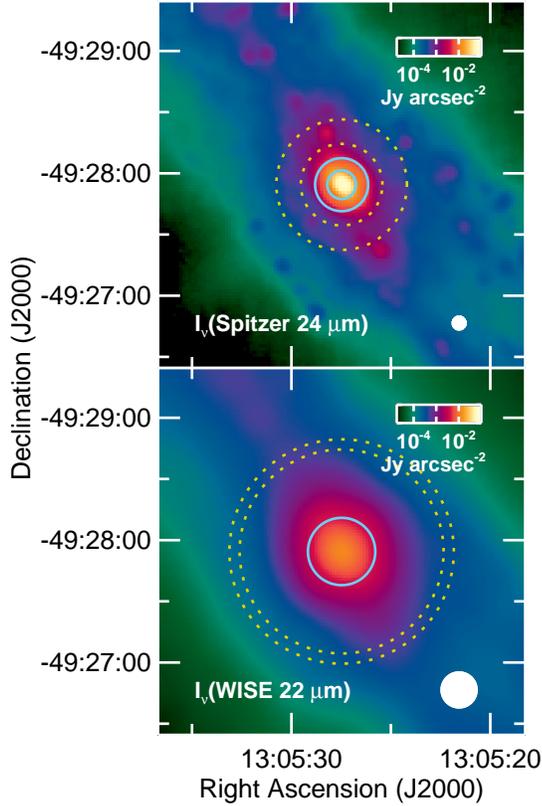,width=7.5cm}
\caption{The {\it Spitzer} 24~$\mu$m image (top) and WISE 22~$\mu$m image (bottom) of the central 3~arcmin of NGC 4945.  These images cover a larger area than the images of the ALMA data (Figures-\ref{f_map_contline}, \ref{f_map_measreg}, \ref{f_map_gh}, and \ref{f_map_te}).  The solid cyan circles show the apertures used to measure the emission from the target.  The two cyan circles in the {\it Spitzer} image have radii of 7 and 13~arcsec, and the cyan circle in the WISE image has a radius of 16.5~arcsec.  The dotted yellow circles show the annuli used for optional background subtraction.  The {\it Spitzer} annulus has radii of 20-32~arcsec, and the WISE annuli have radii of 50-55~arcsec.  See the caption for Figure~\ref{f_map_contline} for other details on the layout.}
\label{f_map_mir}
\end{figure}

Because of the nonlinearity or saturation effects within radii of 3.8~arcsec, we performed measurements in two apertures with radii corresponding to those for which \citet{engelbracht07} give aperture corrections: a circular aperture with a radius of 13~arcsec and an annulus with radii of 7 and 13~arcsec, which excludes the emission from the central pixels.  We used the \citet{engelbracht07} aperture correction for the circular aperture and used their 7 and 13~arcsec data to derive the aperture correction for the annulus.  For both apertures, we made measurements without performing any local background subtraction (and relying solely on the global background subtraction) and with the removal of a local background measured within an annulus with radii of 20 and 32~arcsec (which correspond to an optional background annulus for use with these aperture corrections).  These target and background apertures are shown in the left panel in Figure~\ref{f_map_mir}.

To measure the infrared flux density in the WISE 22~$\mu$m data, we used a single circular measurement aperture of 16.5~arcsec, which corresponds to the standard aperture in the WAPPco photometry system used to produce the WISE All-Sky Data Release Products \citep{cutri13}.  An aperture correction is provided for this measurement aperture, but it is intended for application to images with no diffuse, extended emission.  We performed measurements in the target aperture using just a global background subtraction (based on measurements outside the optical disc) and based on a local background subtraction where the local background is measured within an annulus of 40-50~arcsec, which is at a location where the radial profile of the beam flattens. We derived the aperture correction for this combination of 16.5~arcsec radius target aperture and 40-50~arcsec background aperture using the WISE beams provided by \citet{cutri13}\footnote{The beams are available at http://wise2.ipac.caltech.edu/docs/release/ allsky/expsup/sec4\_4c.html}.

\begin{table*}
\centering
\begin{minipage}{100mm}
\caption{Star formation rates for the central disc measured from mid-infrared data.}
\label{t_mir}
\begin{tabular}{@{}lccccc@{}}
\hline
Band &
    Target &
    Background &
    Aperture &
    Flux &
    SFR \\
&
    Aperture &
    Annulus &
    Correction &
    Density &
    (M$\odot$ yr$^{-1}$)$^a$\\
&
    Radius &
    Radii &
    &
    (Jy) &
    \\
&
    (arcsec) &
    (arcsec) &
    &
    &
    \\
\hline
{\it Spitzer} 24~$\mu$m &
  13 &
  [none] &
  1.17 &
  $9.2\pm0.4$ &
  $0.41$ \\
&
  13 &
  20-32 &
  1.17 &
  $8.9\pm0.4$ &
  $0.39$ \\
&
  7-13 &
  [none] &
  4.15 &
  $12.6\pm0.5$ &
  $0.55$ \\
&
  7-13 &
  20-32 &
  4.28 &
  $11.9\pm0.5$ &
  $0.53$ \\
WISE 22~$\mu$m &
  16.5 &
  [none] &
  1.76 &
  $9.2\pm0.2$ &
  $0.41$ \\
&
  16.5 &
  50-55 &
  1.77 &
  $8.8\pm0.2$ &
  $0.39$ \\
\hline
\end{tabular}
$^a$ The uncertainties in the SFR are 0.20~dex, which is equivalent to a factor of 1.6.
\end{minipage}
\end{table*}

Several equations for converting {\it Spitzer} 24~$\mu$m or WISE 22~$\mu$m flux densities to SFR have been published \citep[see ][ for reviews]{calzetti10, lee13}.  We use
\begin{equation}
  \frac{\mbox{SFR}(24\mu\mbox{m})}{\mbox{M}_\odot\mbox{ yr}^{-1}}=
  7.8\times10^{-10} \frac{\nu L_\nu(24\mu\mbox{m})}{\mbox{L}_\odot}
\label{e_sfr_mir}
\end{equation}
given by \citet{rieke09}, as this is one of the most commonly-used relations.  The relation was derived primarily for application to globally-integrated mid-infrared fluxes where $6\times10^{8}\mbox{L}_\odot\leq L(24\mu\mbox{m})\leq1.3\times10^{10}\mbox{L}_\odot$.  The central region in NGC~4945 falls near or below the low end of this range (depending on how we measure the flux density), but as discussed by \citet{calzetti10} and \citet{lee13}, relations with similar coefficients have been derived for samples with luminosity ranges that bracket the NGC~4945 central starburst.  This equation was derived assuming that the star forming region is optically thick and that all photoionizing light is absorbed by dust, which should be applicable to the centre of NGC~4945.  The results should be reliable to within 0.2 dex, which is equivalent to a factor of 1.6.  

The measured flux densities and SFRs are listed in Table~\ref{t_mir}.  The measurements are relatively unaffected by the inclusion of local background subtraction, probably because the surface brightness of the central disc is substantially higher than the surface brightness of the diffuse emission.  Between the two {\it Spitzer} target apertures used, the 7-13~arcsec radius annulus yielded higher values, although the change in measurements is less than a factor of 2.  This could be consistent with saturation in the central pixels in the {\it Spitzer} data, or it could also be consistent with the inclusion of emission from additional diffuse or fainter star forming regions within the broader measurement apertures or issues with the aperture correction derived for the annulus, which is relatively high.  The WISE flux density measurement are statistically similar to the {\it Spitzer} measurements within the 13~arcsec circular aperture, which implies that any artefacts in the {\it Spitzer} data have a minor affect on the results.  Based on this, we will work with a mid-infrared SFR measurement of 0.4 M$_\odot$ yr$^{-1}$, which corresponds to the measurement in the 13~arcsec aperture in the {\it Spitzer} data and the WISE measurement.

\subsection{Star formation rates from total infrared fluxes}

For calculating a total infrared flux, we combined the {\it Spitzer} 24~$\mu$m measurement with measurements from the {\it Herschel} 70-500~$\mu$m data.   The central source, which we assume to be the approximately the same size as the T region in Figure~\ref{f_map_measreg}, appears unresolved in the 70-500~$\mu$m bands, and the surface brightness in the centre of the galaxy is $\gtsim$30$\times$ higher than the surface brightness within the rest of the optical disc of the galaxy in most of these bands, so we used photometry methods optimized for point sources.  Because of the coarser resolution of the data, however, these measurements may include additional emission from circumnuclear structures that are more extended than the structures seen in the ALMA data, and this could bias the SFR to high values.  Since the optical design and flux calibration of the 70-160~$\mu$m PACS detectors and 250-500~$\mu$m SPIRE detectors differ from each other, we used different techniques to measure the flux densities from these instruments.

\begin{figure*}
\epsfig{file=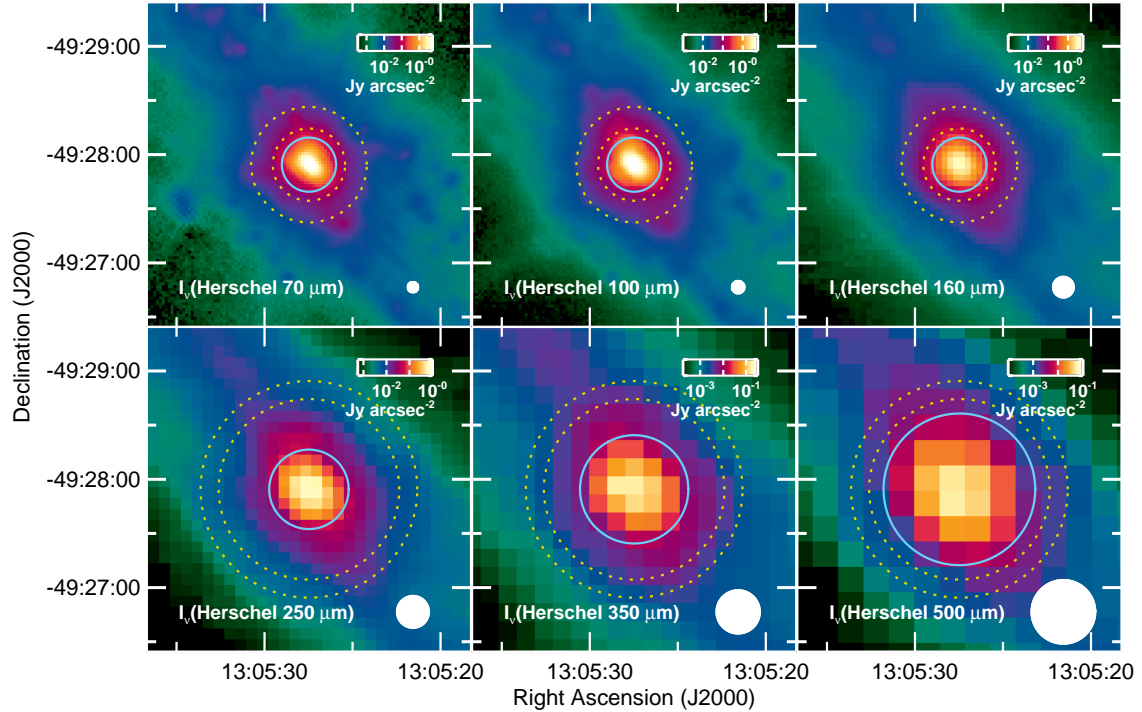}
\caption{The {\it Herschel} 70-500~$\mu$m images of the central 3~arcmin of NGC 4945.  These images cover the same area as the mid-infrared images in Figure~\ref{f_map_mir} and a larger area than the images of the ALMA data (Figures-\ref{f_map_contline}, \ref{f_map_measreg}, \ref{f_map_gh}, and \ref{f_map_te}).  The solid cyan circles in the 70-160~$\mu$m images show the 13~arcsec apertures used to measure the emission from the target, and the dotted yellow circles show the annuli with radii of 20-32~arcsec used for local background subtraction.  The solid cyan circles in the 250-500~$\mu$m images show the regions within which the timeline data were fit with a two-dimensional Gaussian function to obtain the flux density of the central source; the radii are listed in Table~\ref{t_flux_spire}.  The dotted yellow circles show the annuli with radii of 50-60~arcsec used to sample the local background in one set of measurements; the 720-730~arcsec annuli used to sample the global background fall outside these panels.  While the 70-160~$\mu$m images were used for photometry, timeline data were used for photometry in the 250-500~$\mu$m bands, so the 250-500~$\mu$m here are only for display purposes.  See the caption for Figure~\ref{f_map_contline} for other details on the layout.}
\label{f_map_fir}
\end{figure*}

To measure flux densities in the PACS data, we measured the signal in the final images within circular apertures, which we could use because the elongation of the unresolved central disc is not visible in these data.  The PACS documentation provides no specific recommendations for apertures within which to measure flux densities, so we used a target aperture of 13~arcsec and a background annulus with radii of 20-32~arcsec, which matches the preferred sets of target and background apertures used for the {\it Spitzer} 24~$\mu$m data.  We applied aperture corrections calculated by {\sc photApertureCorrectionPointSource} in {\sc HIPE}, and we applied colour corrections based on the midpoint of the range given by \citet{muller11} for modified blackbodies with temperatures of 20-50~K and emissivities scaling as either $\nu^{1}$ or $\nu^{2}$, which should represent the extreme ranges of the dust emission at these wavelengths (and the colour temperatures of the dust are consistent with these ranges).  The measurements and correction factors are listed in Table~\ref{t_flux_pacs}, and the measurement apertures are shown in Figure~\ref{f_map_fir}.

To measure flux densities in the SPIRE data, we fit the timeline data with a two-dimensional circular Gaussian function.  As described by \citet{bendo13}, this method should provide accurate flux densities for unresolved sources like the central starburst in NGC~4945.  To sample the centre of the beams, we selected data within galactocentric radii of 22, 30, and 42~arcsec at 250, 350, and 500~$\mu$m, respectively, as recommended by the SPIRE Data Reduction Guide \citep{spire15}\footnote{This document is available from http://herschel.esac.esa.int/hcss-doc-14.0/print/spire\_drg/spire\_drg.pdf .}.  For sampling the background emission, we experimented with using smaller annuli with radii of 50-60~arcsec (which sample the background close to the central source but may be affected by its beam) and larger annuli with radii of 720-730~arcsec (which fall outside the optical disc of the galaxy); the use of differing background annuli affected the results by $<$1\%.  We applied colour corrections based on the midpoint of the range given by the SPIRE Observers' Handbook for point-like modified blackbodies with temperatures of 15-40~K and emissivities scaling as either $\nu^{-1.5}$ or $\nu^{-2}$, which is a broad temperature and emissivity range that is typical for extragalactic objects.  Uncertainties in the colour correction are based on the extreme values in this range.  Table~\ref{t_flux_spire} lists the photometry results.

\begin{table}
\caption{Flux density measurements for the central source in {\it Herschel}-PACS 70-160~$\mu$m data.}
\label{t_flux_pacs}
\begin{center}
\begin{tabular}{@{}lccc@{}}
\hline
Wavelength &
  Aperture &
  Colour &
  Flux \\
($\mu$m) &
  Correction &
  Correction &
  Density (Jy)\\
\hline
70 &
  1.23 &
  $1.06 \pm 0.09$ &
  $790 \pm 80$ \\
100 &
  1.26 &
  $1.00 \pm 0.02$ &
  $1050 \pm 60$ \\
160 &
  1.44 &
  $1.03 \pm 0.08$ &
  $870 \pm 80$ \\
\hline
\end{tabular}
\end{center}
\end{table}

\begin{table}
\caption{Flux density measurements for the central source in {\it Herschel}-SPIRE 250-500~$\mu$m data.}
\label{t_flux_spire}
\begin{center}
\begin{tabular}{@{}lcccc@{}}
\hline
Wavelength &
  Colour &
  Target &
  Background &
  Flux \\
($\mu$m) &
  Correction &
  Aperture &
  Annulus &
  Density \\
&
  &
  Radius &
  Radii &
  (Jy)\\
&
  &
  (arcsec) &
  (arcsec) &
  \\
\hline
250 &
  $0.955 \pm 0.045$ &
  22 &
  50-60 &
  $305 \pm 19$ \\
&
  &
  &
  720-730 &
  $306 \pm 19$ \\
350 &
  $0.935 \pm 0.035$ &
  30 &
  50-60 &
  $127 \pm 7$ \\
&
  &
  &
  720-730 &
  $127 \pm 7$ \\
500 &
  $0.905 \pm 0.035$ &
  42 &
  50-60 &
  $42 \pm 2$ \\
&
  &
  &
  720-730 &
  $42 \pm 2$ \\
\hline
\end{tabular}
\end{center}
\end{table}

To calculate the total infrared flux, we first interpolated among the logarithms of the {\it Spitzer} 24~$\mu$m flux density (from the 13~arcsec aperture with the local background subtraction) and the 70-500~$\mu$m flux densities listed in Tables~\ref{t_flux_pacs} and \ref{t_flux_spire} with a spline function and then integrate underneath the function, yielding a flux of $2.0\times10^{10}$~L$_\odot$.  Uncertainties in the flux densities affect the total infrared flux by 5\%, but alterations in the interpolation technique, such as using linear interpolation between the points, alters the the total infrared flux by up to 10\%, and up to 10\% additional emission may originate from outside the 24-500~$\mu$m range.  We therefore estimate the uncertainty of the total flux to be $0.2\times10^{10}$~L$_\odot$.

\begin{figure}
\epsfig{file=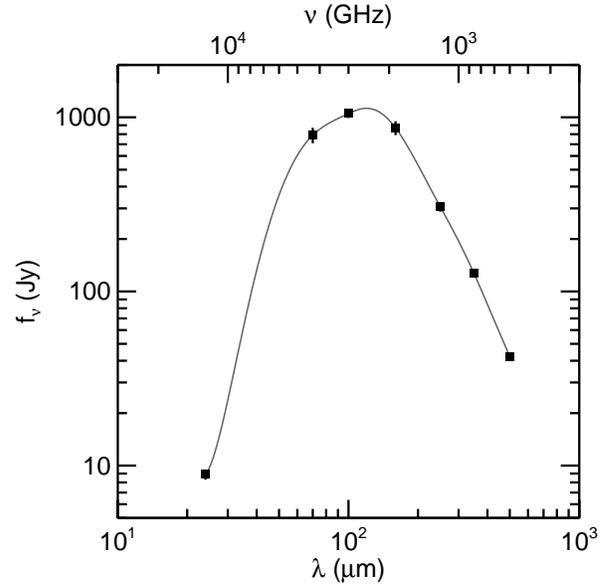}
\caption{The infrared SED for the central source seen in the 24-500~$\mu$m images.  The line is the result of interpolating between the points in logarithm space using a spline function.}
\label{f_irsed}
\end{figure}

The infrared SED of the central source and the spline function are shown in Figure~\ref{f_irsed}.  The SED is very steep between 24 and 70~$\mu$m, which is unusual compared to the SEDs of most galaxies where the dust is assumed to mostly be optically thin (for example, see the model SEDs fit to nearby galaxy SEDs presented by \citet{ciesla14}).  However, it would be expected if the dust is optically thick, as is the case for other nearby infrared-luminous objects \citep[e.g.][]{rangwala11}.  The deviation of the SED from standard dust emission models is why we do not use previously-developed equations that produce a total infrared flux as a sum of fluxes measured in individual wavebands multiplied by scaling terms \citep[see ][ and references therin]{dale14}.  Having said that, if we apply the \citet{galametz13} conversion from the 24, 70, 100, 160, and 250~$\mu$m flux densities to total infrared flux, we get a similar number to what we measure by integrating under the curve, although this flux estimate could change by 20\% if we use one of the \citet{galametz13} conversions that does not include all of these bands.

To convert the total infrared flux to SFR, we use
\begin{equation}
\frac{\mbox{SFR}(\mbox{total infrared})}{\mbox{M}_\odot\mbox{ yr}^{-1}}
  =1.49\times10^{-10}\frac{L(\mbox{total infrared})}{\mbox{L}_\odot}
\label{e_sfr_tir}
\end{equation}
from \citet{murphy11}.  This conversion is based on a stellar population with a Kroupa IMF, in which sense it is comparable to the equations used in Section~\ref{s_sfr_alma} to calculate SFR from ALMA continuum and recombination line data.  Equation~\ref{e_sfr_tir} does not account for stellar rotation, but using results from {\sc Starburst99}, we estimate that the change in the conversion would only be on the order of 10\% if stellar rotation was accounted for in the same way as for the ALMA data.  The final SFR measured for the central source based on the total infrared flux is $3.0 \pm 0.3$~M$_\odot$~yr$^{-1}$.

\subsection{Star formation rates from previously-published radio data}
\label{s_sfr_radio}

\citet{roy10} have published the only other higher-order recombination line emission detections for NGC 4945.  Because the H91$\alpha$ and H92$\alpha$ lines observed by \citet{roy10} are produced at lower frequencies, the line emission could be affected by masing or optical thickness effects, so to derive SFRs, they used models of the line emission rather than a simplified multiplicative conversion.  The best fitting models have a range of SFRs from 2 to 8~M$_\odot$ yr$^{-1}$.

\citet{lenc09} published resolved images of supernova remnants within the central 10~arcsec of the galaxy.  Based on supernovae source counts, sizes, and expansion rates, they express the possible range of SFR as $2.4(v/10^4)<\mbox{SFR}(M>5\mbox{M}_\odot)<370$~M$_\odot$ yr$^{-1}$, where $v$ is the radial expansion velocity of supernova remnants in km~s$^{-1}$.  For a Kroupa IMF, SFR$(M\geq$ 5 M$_\odot$) is approximately equivalent to 0.32 SFR for all stellar masses.  If we use $v$=$10^4$~km~s$^{-1}$ as an approximation, then we have $7.5<\mbox{SFR}<1170$~M$_\odot$ yr$^{-1}$.

We can also convert the radio continuum emission to SFR.  \citet{murphy11} gives a relation between synchrotron continuum emission and SFR that can be expressed as
\begin{equation}
\begin{split}
\frac{\mbox{SFR (synchrotron)}}{\mbox{M}_\odot\mbox{ yr}^{-1}}=
  \ \ \ \ \ \ \ \ \ \ \ \ \ \ \ \ \ \ \ \ \ \ \ \ \ \ \ \ \ \ \ \ \ \ \ \ \ \ \ \ \ \ \ \ \ \ \ \ \ \ \ \ \ \ \ \ \ \ \ \ \\
  0.0797
  \left[\frac{\nu}{\mbox{GHz}}\right]^{-\alpha}
  \left[\frac{D}{\mbox{ Mpc}}\right]^{2}
  \left[\frac{f_\nu(\mbox{cont})}{\mbox{Jy}}\right]
\end{split}
\label{e_sfr_sync}
\end{equation}
where $\alpha$ is the spectral index of -1.40 from the fit to the SED performed in Section~\ref{s_sed}.  This was derived for a Kroupa IMF.  \citet{murphy11} indicate that the coefficient within this equation is dependent upon the IMF used in the calculation and that the range of previously-published coefficients has varied by a factor of $\sim$7.  Since the SFR calculations we performed in Section~\ref{s_sfr_alma} are also based on the Kroupa IMF, the results from Section~\ref{s_sfr_alma} should be comparable to what is obtained using the above equation, although some caution is still needed when interpreting the results.

\citet{condon92} and \citet{murphy11} give equations for converting radio emission to SFR that are based on assuming that the total emission observed at any frequency comprises a component representing synchrotron emission and a component representing free-free emission.  If we take this approach using Equations~\ref{e_sfr_freefree}\footnote{\citet{murphy11} also give an expression relating free-free emission to SFR, but the relation is based on an older expression for the flux density from free-free emission published by \citet{rubin68}.  The relation we used, which is derived from the \citet{draine11} expression for free-free emission, incorporates a newer estimate of the gaunt factor and newer versions of effective recombination coefficients.} and \ref{e_sfr_sync}, we obtain
\begin{equation}
\begin{split}
\frac{\mbox{SFR (radio)}}{\mbox{M}_\odot\mbox{ yr}^{-1}}=
  \left[ 
  12.5 \left[ \frac{\nu}{\mbox{GHz}} \right]^{\alpha}
  + 1.05\times10^{-11} g_{ff} 
  \vphantom{\left[\frac{T_e}{\mbox{~K}}\right]^{-0.5}}\right.
  \ \ \ \ \ \ \ \ \ \ \ \\
  \left.
  \times
  \left[\frac{\alpha_B}{\mbox{ cm}^3\mbox{ s}^{-1}}\right]^{-1}
  \left[\frac{T_e}{\mbox{~K}}\right]^{-0.5} 
  \right]^{-1}
  \left[\frac{D}{\mbox{ Mpc}}\right]^{2}
  \left[\frac{f_\nu(\mbox{cont})}{\mbox{Jy}}\right].
\end{split}
\label{e_sfr_radio}
\end{equation}
This expression effectively includes assumptions about the relative contributions of free-free and synchrotron emission to the overall continuum emission.  Moreover, it effectively relies on the assumption that the SFR has not changed between the time when the progenitors of the supernovae formed $\sim$3-40~Myr ago and the time when the currently-observed photoionizing stars formed in the past $\sim$5~Myr.  Variations in SFR over time would lead to ratios of synchrotron emission to free-free emission that differ from the assumptions used when deriving Equation~\ref{e_sfr_radio}, and the resulting SFR would potentially fall anywhere within the range of SFRs within the past 40~Myr.  Unfortunately, the ratios predicted by this equation do not match our SED decomposition results from Section~\ref{s_sed}, which causes the SFR from this equation to deviate from the results from Section~\ref{s_sfr_alma}.  At 4.8~GHz, this equation would predict that the free-free and synchrotron emission each produce approximately half of the total emission, but the results from our SED fits show that $\sim$80\% of the total emission is synchrotron emission.  At 85.69~GHz, the equation would predict that $\sim$96\% of the total emission is from free-free emission, which seems slightly high compared to our SED fitting result.

\begin{table}
\caption{SFR from Equation~\ref{e_sfr_radio} based on radio continuum measurements from the literature.}
\label{t_sfr_radio}
\begin{center}
\begin{tabular}{@{}lcc@{}}
\hline
Frequency &
  Flux &
  SFR \\
(GHz) &
  Density (Jy)$^a$ &
  (M$_\odot$ yr$^{-1}$)\\
\hline
4.5-5.0 &
  $2.71 \pm 0.34$ &
  $13.5 \pm  1.5$ \\
8.0-9.0 &
  $1.27 \pm 0.17$ &
  $ 8.9 \pm  1.2$ \\
17-23 &
  $0.78 \pm 0.08$ &
  $ 7.8 \pm  0.8$ \\
85.69 (total emission; Eq. \ref{e_sfr_radio}) &
  $0.36 \pm 0.02$ &
  $5.1 \pm  0.3$ \\
85.69 (free-free emission; Eq. \ref{e_sfr_freefree}) &
  $0.31 \pm 0.04$ &
  $4.42 \pm 0.49$ \\
\hline
\end{tabular}
\end{center}
$^a$ Except for the 85.69~GHz data, these numbers represent the mean and standard deviation for the flux densities in Table~\ref{t_sed} that fall within the given frequency range.\\
\end{table}

Despite these issues, we still want to assess this method of converting radio emission to SFR that was suggested by \citet{condon92} and \citet{murphy11}, so we applied Equation~\ref{e_sfr_radio} to the average flux densities for measurements between 4.8-5, 8-9, and 17-23~GHz listed in Table~\ref{t_sed}.  These SFR data are listed in Table~\ref{t_sfr_radio}.  The numbers decrease systematically as frequency increases.  This could be related to issues with the inaccurate estimates of the relative contribution of synchrotron emission to the SED at these frequencies, or the synchrotron component in Equation~\ref{e_sfr_radio} may not be scaled correctly.  We also applied Equation~\ref{e_sfr_radio} to the total 85.69~GHz flux density measured within the T region in Figure~\ref{f_map_measreg}.  The SFR computed from Equation~\ref{e_sfr_radio} is $\sim$15\% higher than what was calculated from Equation~\ref{e_sfr_freefree} for just the free-free component, which is a reasonably good match considering the various assumptions applied in Equation~\ref{e_sfr_radio}.  These results probably match because most of the emission at 85.69~GHz is from free-free emission, so inaccuracies in the relative fraction of emission from synchrotron emission or the conversion of synchrotron emission to SFR do not cause the results to differ significantly from each other.

\citet{lenc09} calculated a SFR based on flux densities integrated over 8.64-23~GHz using a conversion equation published by \citet{haarsma00} (which were an alternate form of the equations published by \citet{condon92}).  Their resulting SFR is expressed as $(14.4\pm1.4)(Q/8.8)$~M$_\odot$~yr$^{-1}$, where $Q$ is the ratio of the total mass of stars to the mass of stars above 5 M$_\odot$ (and which should not be confused with the term $Q$ we use for the photoionizing photon production rate).  \citet{lenc09} state that $Q$ is 8.8 for an IMF with a power law index of -2.5 (which they incorrectly identify as the index for a \citet{salpeter55} IMF).  For a Kroupa IMF, $Q$ is equal to 3.2.  When applied to the expression for the SFR from \citet{lenc09}, this gives $5.2 \pm 0.5$M$_\odot$~yr$^{-1}$, which is slightly below the value that was derived using Equation~\ref{e_sfr_radio} applied to the data from \citet{lenc09} and other data from the literature at these frequencies.

\subsection{Discussion}

Table~\ref{t_sfrsummary} presents a summary of all of the SFR measurements discussed in Sections~\ref{s_sfr_alma}-\ref{s_sfr_radio}.  We first discuss how the ALMA SFR measurements compare to the SFR measurements from total infrared and radio data, as all of these measurements differ by less than a factor of 2.  We then will focus on comparisons to the mid-infrared SFR measurements, which are $\sim$10$\times$ lower than other SFR measurements.

\subsubsection{Comparison of ALMA and total infrared SFR measurements}

Although the SFR from the the total infrared flux is $\sim$25\% lower than that measured from the ALMA data, the correspondence is still reasonably good, especially considering the limitations of the infrared photometry, the uncertainties in the total infrared flux calculation, and the various assumptions behind the conversion of the ALMA recombination line and continuum emission as well as the conversion of the total infrared flux to SFR.   Variations in the star formation history could affect the match in the SFR from the ALMA and total infrared data, as the recombination line and free-free emission trace photoionizing stars that have formed in the past $\ltsim$5~Myr, whereas the total infrared flux is connected to the bolometric luminosity from the total stellar population including stars older than 5~Myr.  

Notably, the total infrared flux does not yield a SFR higher than the SFR from the ALMA data.  As stated in the introduction, the AGN is expected to be heavily obscured by interstellar dust, as no signature of the AGN emission is seen in mid-infrared spectral line data.  If the AGN was a significant dust heating source, then SFR measurements from the total infrared flux should be high compared to the ALMA data.  This indicates that the starburst is the dominant heating source for infrared emission seen from the centre of NGC~4945.

\subsubsection{Comparison of ALMA and radio SFR measurements}

The SFRs from the ALMA data fall within the range of potential SFR values given by \citet{roy10}, which is the only other analysis based on recombination line data.  However, the SFRs from the ALMA data are typically a factor of $\sim$2 lower than what would be derived from radio data.  This could arise for two reasons.  First, the scaling term in the conversion between synchrotron emission and SFR may be inaccurate or inapplicable to our data for multiple reasons.  The scaling term itself may be highly uncertain, as discussed by \citet{schmitt06} and \citet{murphy11}.  Additionally, the relative contribution of synchrotron emission to the SED may not be accurately characterized by equations like Equation~\ref{e_sfr_radio} that are used to convert radio continuum emission to SFR.  Also, the term may be better scaled for situations where the spectral index of the synchrotron emission is closer to -0.8.  In our data, we have measured a steeper spectral index, which may be the result of our choice of frequencies.  The spectral slope is similar to what is derived by \citet{peel11} for nearby galaxies, and they also excluded lower-frequency data in their SED fitting analysis.  

Aside from normalization issues, however, it is possible that the direct supernova observations and integrated lower-frequency radio continuum data yield higher SFRs than recombination line and ALMA free-free continuum measurements because the measurements are tracing SFR at different epochs.  As stated above, recombination line and free-free emission will trace stars formed within the past $\sim$5~Myr.  Supernovae and the synchrotron emission associated with supernovae will be directly proportional to the SFR that occurred $\sim$3-40~Myr ago.  If the SFR has decreased by a factor of 2 within the past few Myr, it could cause the discrepancy between the ALMA SFR measurements and the SFR values from either the radio continuum data or the direct supernova observations.

\begin{table}
\caption{Summary of SFR measurements for the central starburst in NGC 4945.}
\label{t_sfrsummary}
\begin{center}
\begin{tabular}{@{}lc@{}}
\hline
Measurement &
  SFR \\
&
  (M$_\odot$ yr$^{-1}$)\\
\hline
85.69 GHz free-free (ALMA)$^a$ &
  $4.42 \pm 0.49$ \\
H42$\alpha$ (ALMA)$^a$ &
  $4.29 \pm 0.07$ \\
Mid-infrared ({\it Spitzer} and WISE) &
  0.4$^b$\\
Total infrared ({\it Spitzer} and {\it Herschel}) &
  $3.0 \pm 0.3$ \\
4.5-5.0~GHz continuum (multiple published measurements) &
  $12.4 \pm  1.4$ \\
8.0-9.0~GHz continuum (multiple published measurements) &
  $ 8.2 \pm  1.1$ \\
17-23~GHz continuum (multiple published measurements) &
  $ 7.4 \pm  0.7$ \\
H91$\alpha$, H92$\alpha$ \citep{roy10} &
  2 - 8 \\
Radio supernovae analysis \citep{lenc09} &
  7.5 - 1170\\
\hline
\end{tabular}
\end{center}
$^a$ The ALMA SFRs used in this comparison are integrated over the entire central disc (the T region in Figure~\ref{f_map_measreg}.\\
$^b$ The uncertainties in the mid-infrared SFR are 0.20~dex, which is equivalent to a factor of 1.6.
\end{table}

\subsubsection{Comparisons of all SFR measurements to mid-infrared SFR measurements}

The resulting SFR from the mid-infrared data is $\sim10\times$ lower than the SFR from the ALMA data, from the total infrared flux, or from any radio data published in the literature.  If anything, it was expected that the mid-infrared fluxes should yield high SFR measurements because they should also include emission from dust heated by the AGN.  Given the agreement (within a factor of $\sim$2) of the other SFR measurements, the most likely explanation is that the SFR from the mid-infrared flux density is abnormally low compared to the actual SFR.  

It is unlikely that the mid-infrared flux densities are measured incorrectly.  While the {\it Spitzer} data could have been affected by detector saturation effects, the fact that we were able to measure the same SFR in the WISE data would indicate that such effects are unlikely to be a cause for the flux density mismatch.  If we corrected the central flux density of the {\it Spitzer} 24~$\mu$m or WISE 22~$\mu$m data so that it corresponded to a SFR of $\sim$4~M$_\odot$~yr$^{-1}$, the global flux densities at these wavelengths would need to be increased by $\sim$80~Jy or a factor of $\sim$3.  The {\it Spitzer} and WISE data show no signs of any instrumental effects that could cause such discrepancies, and if the {\it Spitzer} or WISE flux densities were corrected by this amount, the results would be discrepant with the IRAS 25~$\mu$m flux densities.

The centre of NGC~4945 may be a source where the mid-infrared flux yields an underestimate of the SFR because the relation between the 24~$\mu$m and total infrared flux is atypical.  The conversion between {\it Spitzer} 24~$\mu$m emission and SFR derived by \citet{rieke09} is based upon the assumption that the ratio of 24~$\mu$m flux (expressed as $\nu f_\nu$) to total infrared flux should be 0.158.  In the case of the centre of NGC~4945, the ratio is $\sim$0.02.  Two phenomena could suppress the mid-infrared emission relative to the total infrared flux in the centre of NGC~4945.  

First, based on empirical analyses, the relation between 24~$\mu$m and total infrared flux is expected to become nonlinear at high infrared surface brightnesses.  \citet{rieke09} state that such a nonlinearity should be seen at $>10^{11}$ L$_\odot$, but this is based on globally-integrated flux densities, whereas the point at which the relation becomes nonlinear should depend primarily on the intensity of the illuminating radiation field.  When the dust emission comes from a very compact source like the centre of NGC 4945, it is possible that the relation becomes nonlinear at luminosities lower than $10^{11}$ L$_\odot$.  However, in the sample of galaxies studied by \citet{rieke09}, the nonlinearity effects are expected to be relatively small and not on the order of a factor of 10.  

It is more likely that the mid-infrared emission is suppressed because the central starburst is optically thick in mid-infrared bands.  This has been seen in other compact luminous sources \citep[e.g.][]{rangwala11}.  The conversion from 24~$\mu$m emission to SFR clearly relies upon the assumption that the dust emission is optically thin.  If the dust is not optically thin in the mid-infrared, the emission in this waveband will be suppressed relative to longer wavelengths, and the resulting SFR will be lower.  The ratio of the mid-infrared SFR (0.4~M$_\odot$ yr$^{-1}$) to the average of the ALMA SFRs (4.35~M$_\odot$ yr$^{-1}$) is 0.092.  Assuming the 24~$\mu$m flux density is suppressed by this factor, the dust attenuation can be expressed as $A_{24\mu m}$ in magnitude units as 2.6.  Applying the \citet{draine03} renormalized versions of the \citet{weingartner01} extinction curve for $R_V$=5.5, which appears to replicate mid-infrared dust extinction measurements within very dusty Milky Way regions \citep[see ][ and references therein]{wang14}, this $A_{24\mu m}$ is equivalent to $A_V$=150 and $A_K$=17.  This high extinction would be consistent with the $>$50 $A_V$ values estimated by \citet{brock88}\footnote{Although our $A_V$ is similar to the result from \citet{brock88}, they measure a 100~$\mu$m flux density of 705 Jy for the nucleus, which is lower than our measurement of $1050\pm60$~Jy.}, \citet{bergman92}, and \citet{perezbeaupuits11}.

Since $\sim$30\% of the global mid-infrared emission from NGC~4945 originates from the central starburst, it is readily apparent that the globally integrated mid-infrared flux density will yield an underestimate of the global star formation rate for this galaxy.  If we apply Equation~\ref{e_sfr_mir} to the globally-integrated {\it Spitzer} 24~$\mu$m flux density of 31.9~Jy, we obtain a SFR of 1.4~M$_\odot$~yr$^{-1}$.  If we correct the nuclear SFR from 0.4 to 4~M$_\odot$~yr$^{-1}$, the global SFR becomes 5~M$_\odot$~yr$^{-1}$.  This is a change of a factor of $\sim$3.5.

These results have major implications for measuring the star formation rate within galaxies with compact, dusty starbursts similar to the one in NGC~4945.  While the central starburst in NGC~4945 is unusual compared to the nuclei of most other galaxies within 10 Mpc of the Milky Way Galaxy, it is similar in intensity circumnuclear starbursts in galaxies like M82, M83, and NGC 253, and it should be representative of the central starbursts seen in many luminous infrared galaxies (LIRGs; $10^{11}\mbox{L}_\odot<L(\mbox{total infrared})<10^{12}\mbox{L}_\odot$) and ultraluminous infrared galaxies (ULIRGs; $L(\mbox{total infrared})>10^{12}\mbox{L}_\odot$).  In NGC~4945, 25-75\% of the emission in any infrared waveband originates from the central region, which is similar to what is seen in many of these other galaxies \citep[e.g.][]{diazsantos10}.  If mid-infrared emission is used by itself to measure global SFRs in these classes of objects, as is very commonly done with {\it Spitzer} or WISE data and as could be done with the {\it James Webb} Space Telescope, the resulting measurements could be significantly biased downwards. 

In addition to the obvious issues with measuring SFR, the results here also have implications for modelling dust emission from galaxies with compact central starbursts.  When applying dust emission models or radiative transfer models to the infrared SEDs of galaxies, it is important to use models that account for not only the high opacities in the mid-infrared but also the shift in dust emission to longer wavelengths.

\section{Conclusions}
\label{s_conclu}

We have presented here ALMA observations of 85.69~GHz continuum emission and H42$\alpha$ line emission from the centre of NGC~4945.  These data are one of only a small number of currently-existing ALMA data that include the detection of recombination line emission from an extragalactic source, and our analysis is one of the earliest comparisons of SFR measurements from ALMA data with SFR measurements from infrared data.  In summary, we have obtained the following results:

\vspace{0.625em}

\setlength{\leftskip}{1em}

\noindent $\bullet$ The 85.69~GHz continuum and H42$\alpha$ line emission originates from a structure that can be modelled as an exponential disc with a scale length of $\sim$2.1~arcsec ($\sim$40~pc).  The spatial extent of the emission as well as the absence of any enhancement in the centre as well as the absence of any broad line emission suggest that the emission originates primarily from photoionized gas associated with the circumnuclear starburst and not from the AGN.

\vspace{0.625em}

\noindent $\bullet$ The SED for the central source implies that 84\%$pm$10\% of the 85.69~GHz continuum emission from the central disc originates from free-free emission.

\vspace{0.625em}

\noindent $\bullet$ The $T_e$ for the central star forming disc based on the ratio of the H42$\alpha$ line emission to 85.69~GHz free-free emission is $5400\pm600$~K.  This is similar to what is measured near the centre of the Milky Way.  These results also imply that the AGN contributes $\ltsim$10\% of the total continuum emission from the central disc.

\vspace{0.625em}

\noindent $\bullet$ The SFR for the central source derived from both the 85.69~GHz continuum and H42$\alpha$ line emission is $4.35\pm0.25$~M$_\odot$~yr$^{-1}$.  This is comparable to what we obtain using the total infrared flux, and it is consistent with the range of SFR values estimated from previous radio recombination line measurements.

\vspace{0.625em}

\noindent $\bullet$ The SFR measurements from either previously-published radio continuum data or from radio observations of supernovae are a factor of $\sim$2 higher than what is obtained from the ALMA data.  This is potentially related to a combination of calibration issues with the estimates of the SFR based on the radio data or changes in the SFR between 3-40~Myr ago and the present.

\vspace{0.625em}

\noindent $\bullet$ The SFR measurements from {\it Spitzer} 24~$\mu$m and WISE 22~$\mu$m data are $\sim$10$\times$ lower than the measurements based on the ALMA data as well as the measurement based on the total infrared flux or the measurements based on radio data.  This probably occurs because the central starburst is optically thick at mid-infrared wavelengths, which is a condition where the conversion between mid-infrared flux and SFR should no longer yield reliable results.

\setlength{\leftskip}{0pt}

\vspace{0.625em}

This analysis not only demonstrates how effective ALMA can be in terms of studying star formation in the centres of starburst galaxies but also how such ALMA observations can be used to cross-check SFR measurements from both infrared and radio data.  Mid-infrared flux has been favoured for use as a star formation tracer because of how well the emission has been correlated with other ultraviolet, optical, and near-infrared star formation tracers and because it had been assumed that mid-infrared emission is not affected by the same dust extinction effects as star formation tracers at shorter wavelength.  The results here demonstrate that mid-infrared fluxes may not be reliable star formation tracers in compact starbursts.  Additional ALMA observations of star forming regions in other nearby galaxies should be used to explore the reliability of infrared emission as a star formation metric for such dusty systems.

\section*{Acknowledgments}

We thank the reviewer for the helpful comments on this paper.  GJB and GAF acknowledge support from STFC Grant ST/M000982/1.  CD has received funding from the European Research Council under the European Union’s Seventh Framework Programme (FP7/2007–2013)/ERC grant agreement no. 307209.  CD also acknowledges support from an STFC Consolidated Grant (no. ST/L000768/1).  AK acknowledges support by the Collaborative Research Council 956, sub-project A1, funded by the Deutsche Forschungsgemeinschaft (DFG).  This paper makes use of the following ALMA data: ADS/JAO.ALMA\#2012.1.00912.S.  ALMA is a partnership of ESO (representing its member states), NSF (USA) and NINS (Japan), together with NRC (Canada), NSC and ASIAA (Taiwan), and KASI (Republic of Korea), in cooperation with the Republic of Chile. The Joint ALMA Observatory is operated by ESO, AUI/NRAO and NAOJ.  This research has made use of the NASA/ IPAC Infrared Science Archive, which is operated by the Jet Propulsion Laboratory, California Institute of Technology, under contract with the National Aeronautics and Space Administration.  This publication makes use of data products from the Wide-field Infrared Survey Explorer, which is a joint project of the University of California, Los Angeles, and the Jet Propulsion Laboratory/California Institute of Technology, and NEOWISE, which is a project of the Jet Propulsion Laboratory/California Institute of Technology. WISE and NEOWISE are funded by the National Aeronautics and Space Administration.

{}

\label{lastpage}

\end{document}